\documentclass[preprint,prc,a4paper,tightenlines,nofootinbib,superscriptaddress]{revtex4}

\usepackage{amssymb}
\usepackage{graphicx}
\usepackage{bm} % for bold greek letters e.g. \bm{\tau}
\usepackage[active]{srcltx}

\usepackage{amsmath}
\usepackage{mathtools}
\usepackage{epsfig}
\usepackage{graphics}
\def\palka{\hspace{-7.5pt}/\hspace{2pt}}
\newcommand{\mpi}{\ensuremath{m_\pi}}   % pion mass
\newcommand{\mn}{\ensuremath{m_{N}}}   % nucleon mass
\newcommand{\mN}{\ensuremath{m_{N}}}   % nucleon mass
\newcommand{\ga}{\ensuremath{g_{\!A}}} % NNpi coupling constant
\newcommand{\gA}{\ensuremath{g_{\!A}}} % NNpi coupling constant
\newcommand{\fpi}{\ensuremath{f_{\!\pi}}} % pion decay constant 92 MeV
\newcommand{\NNLO}{NNLO} % please, use \NNLO{} to avoid problems with spaces
\newcommand{\taup}{\tau_{\!+}}

\newcommand{\taux}{\tau_{\!\times}}

\newcommand{\boldtau}{\bm{\tau}}

\newcommand{\ipipi}{{I_{\pi \pi}}}
\newcommand{\jpid}{{J_{\pi  \Delta }}}
\newcommand{\jpipid}{{J_{\pi \pi \Delta}}}
\newcommand{\jpipind}{{J_{\pi \pi N \Delta}}}

\newcommand{\intI}{\ipipi}
\newcommand{\intJ}{\frac{\jpid}{\delta}}
\newcommand{\intT}{\delta \jpipid}
\newcommand{\intC}[1]{\frac{#1}{(4\pi)^2}}

\newcommand{\gpind}{g_{\pi N \Delta}}

\begin{document}

\title{Threshold pion production in proton-proton collisions\\ at NNLO in chiral EFT}

\author{V.~Baru}
\affiliation{\footnotesize Institut f\"ur Theoretische Physik II, Ruhr-Universit\"at Bochum, D-44780 Bochum, Germany}
\affiliation{\footnotesize Institute for Theoretical and Experimental Physics,
 117218, B.~Cheremushkinskaya 25, Moscow, Russia}

\author{E.~Epelbaum }
\affiliation{\footnotesize Institut f\"ur Theoretische Physik II, Ruhr-Universit\"at Bochum, D-44780 Bochum, Germany}

\author{A.~A.~Filin}
\affiliation{\footnotesize Institut f\"ur Theoretische Physik II, Ruhr-Universit\"at Bochum, D-44780 Bochum, Germany}

\author{C.~Hanhart}
\affiliation{\footnotesize Institut f\"{u}r Kernphysik,  (Theorie) and J\"ulich Center for Hadron Physics,
 Forschungszentrum J\"ulich,  D-52425 J\"{u}lich, Germany and\\
Institute for Advanced Simulation, Forschungszentrum J\"ulich,  D-52425 J\"{u}lich, Germany}

\author{H.~Krebs}
\affiliation{\footnotesize Institut f\"ur Theoretische Physik II, Ruhr-Universit\"at Bochum, D-44780 Bochum, Germany}

\author{F.~Myhrer}
\affiliation{\footnotesize Department of Physics and Astronomy, University of South Carolina,
Columbia, SC 29208, USA}

\today

\begin{abstract}
The reaction $NN \to NN \pi$ offers a good testing ground for
chiral effective field theory at intermediate energies. It challenges our
understanding of the first inelastic channel in
nucleon-nucleon scattering and of the charge-symmetry breaking pattern
in hadronic reactions.
In our previous studies,  
we presented a  complete calculation of the pion-production operator for s-wave pions
 up-to-and-including next-to-next-to-leading order (NNLO) in the formulation
 of chiral effective field theory,
which includes pions, nucleons and $\Delta(1232)$ degrees of freedom. 
In this paper we calculate  the near threshold cross section  for the $pp \to d \pi^{+}$ reaction by performing
the convolution of the
obtained operators with  nuclear wave functions based on  modern phenomenological
and chiral potentials. 
The available chiral $NN$  wave functions are constructed with
a cutoff comparable with  the momentum transfer scale  inherent in pion production  reactions.
Hence, a significant portion of the dynamical
intermediate-range physics is thereby cut off by them. On the other hand,   the  NNLO amplitudes
evaluated with phenomenological wave functions appear to be largely  independent of the $NN$ model used and
give corrections to the dominant leading order  contributions as  
expected from dimensional analysis.
 The result gives support to the counting scheme used  to classify the pion production operators,
which is a precondition for a reliable investigation of  the chirally suppressed neutral pion production.
The explicit inclusion  of the $\Delta(1232)$ is found to be important but smaller than expected due to cancellations.

\end{abstract}

\maketitle

%%%%%%%%%%%%%%%%%%%%%%%%%%%%%%%%%%%%%%%%%%%%%%%%%%
\section{Introduction}
%%%%%%%%%%%%%%%%%%%%%%%%%%%%%%%%%%%%%%%%%%%%%%%%%%

The investigation of near-threshold pion production in proton-proton collisions
is important in order to
gain insights into the dynamics involved in
these first inelastic nucleon-nucleon ($NN$) scattering  channels.
The pion production reaction requires a relatively large
three-momentum transfer and tests the applicability of
chiral effective field theory (EFT) at intermediate energies.
A good theoretical understanding of the isospin-invariant channels is
an important prerequisite for investigations of charge symmetry
breaking (CSB) in few-nucleon reactions such as e.g.~the process  $pn \to d
\pi^0$, see recent review
articles~\cite{Hanhart:2003pg,Baru:2013zpa} and references therein.
The production of pions from two nucleons contributes as a building
block to many few-nucleon processes~\cite{Hanhart:2000gp,Baru:2009fm} and provides the dominant
short-range mechanism in the three-nucleon force~\cite{vanKolck:1994yi,Epelbaum:2002vt}.
It can be investigated experimentally and theoretically
in inelastic nucleon-nucleon reactions, such as
$pp \to d \pi^+$ and $pp \to pp \pi^0$.
When the pioneering work of Koltun and Reitan~\cite{Koltun:1965yk}
was confronted with the high-quality data of Meyer {\it et al.}~\cite{meyer1992},
it was realized that the pion production dynamics was not well understood.
Especially, the reaction
$pp \to pp \pi^0$  appeared to be the most puzzling process.
The experimentally measured  cross section for this process~\cite{meyer1992}
was found to be about $\sim 5$ times larger than the prediction of
Ref.~\cite{Koltun:1965yk}.

Low-energy pion dynamics is governed by the
chiral symmetry of strong interactions and its breaking pattern.
It thus can be naturally addressed in the framework of
chiral  EFT, see
Refs.~\cite{Cohen:1995cc,Park:1995ku} for pioneering studies along
this line and
Refs.~\cite{Sato:1997ps,Hanhart:2002bu,Lensky:2005jc,Filin:2012za,Filin:2013uma}
for more recent applications of chiral EFT to the pion-production reaction\footnote{For an overview of the phenomenological approaches the interested reader is referred
to the review articles Refs.~\cite{Hanhart:2003pg,Baru:2013zpa}}.
As mentioned, the measured near-threshold cross section for the neutral pion production
channel is
suppressed by almost an order of magnitude compared to
charged pion-production channels.
This suppression is naturally explained in chiral EFT
\cite{Hanhart:2003pg,Baru:2013zpa}, where one finds
the leading-order amplitude to $pp \to pp \pi^0$ to be numerically
very small because the dominant isovector  Weinberg-Tomozawa operator does not contribute.
A  quantitative understanding of neutral pion
production therefore requires the inclusion of higher-order
corrections~\cite{Cohen:1995cc,Park:1995ku,Filin:2012za,Filin:2013uma}.

The reaction
$pn \to d \pi^0$ is an
important channel for the study  of CSB
in few-nucleon strong interactions~\cite{Miller:2006tv}.
Specifically, the experimentally measured differential-cross-section
asymmetry~\cite{Opper:2003sb} in this reaction
has been computed using chiral EFT and was used to extract
the strong-interaction contribution to the
neutron-proton mass difference~\cite{vanKolck:2000ip,Bolton:2009rq,Filin:2009yh}.
However, in order to  extract reliably CSB observables,
it is imperative to have an accurate description
of the dominant isospin-symmetric amplitude and  to
ensure that the expansion of chiral EFT converges.
To examine the
convergence of chiral EFT,
the $pp \to d \pi^{+}$ channel is a preferable reaction
since:
(i) unlike $pp \to pp \pi^0$, there is no suppressions of leading-order (LO) contribution
 in this charged pion production
channel~\cite{Koltun:1965yk,Lensky:2005jc,Filin:2012za,Filin:2013uma},
and
(ii) precise experimental data from hadronic atom measurements are
available~\cite{Strauch:2010rm,Strauch:2010vu}.
In this work we summarize the various contributions to
the pion production operator, which were evaluated up to
next-to-next-to-leading order (NNLO) in
Refs.~\cite{Filin:2012za,Filin:2013uma}, and calculate
   the $pp \to d \pi^{+}$ cross section near threshold  in order to test the convergence of
chiral expansion of chiral EFT.\@

Our paper is organized as follows.
In section~\ref{sec:observ} we define the relations between observables and amplitudes in
$pp \to d \pi^+$ channel.
In section~\ref{sec:formalismandpc} we discuss the methods used to calculate
the amplitudes for $NN \to NN \pi$ processes in the chiral EFT framework.
In particular, we outline the counting scheme used to calculate pion production operators
where the intermediate momentum transfer is larger than the pion mass.
In section~\ref{sec:operators} we discuss  the complete NNLO operators for
s-wave pion production given in the Appendix~\ref{app:operators}.
In Sections~\ref{sec:convphenom} and~\ref{sec:convchiral} we perform a convolution of NNLO
pion production operators with phenomenological
and chiral $NN$ wave functions, calculate observables, and compare them with experimental data.
We summarize our findings in section~\ref{sec:concl}.

%%%%%%%%%%%%%%%%%%%%%%%%%%%%%%%%%%%%%%%%%%%%%%%%%%%
\section{The $pp \to d \pi^+$ cross section near threshold}
\label{sec:observ}
%%%%%%%%%%%%%%%%%%%%%%%%%%%%%%%%%%%%%%%%%%%%%%%%%%%
% In this section we provide expression for the observables in $pp \to d \pi^+$ channel.

The differential cross section of the $pp \to d \pi^+$ reaction in the
center-of-mass system (CMS)
is expressed in terms of the transition amplitude as:
\begin{eqnarray}
  \label{eq:dcs}
  \frac{d \sigma}{d \Omega} = \frac{1}{64 \pi^2} \frac{|\vec q_\pi|}{|\vec p_{i}|\, s}
   |\overline{M}|^2,
\end{eqnarray}
where $\vec q_\pi$ is the outgoing pion momentum, $\vec p_{i}$
is the incoming proton momentum,
$\sqrt{s}$ is the total energy in CMS, and $|\overline{M}|^2$ is the
square of the transition amplitude averaged over the spins of the initial protons
and summed over the outgoing deuteron polarizations,
\begin{eqnarray}
  |\overline{M}|^2 = \frac{1}{4} \sum \limits_{\lambda_1 \lambda_2}
   \sum \limits_{\epsilon_d}
  |M_{pp\to d \pi^+}(\lambda_1, \lambda_2, \vec \epsilon_d, \vec p_{i}, \vec q_\pi)|^2 \, .
  \label{eq:avgampl}
\end{eqnarray}
Here  
$\lambda_1$ and $\lambda_2$ are spin projections for each proton,
and $\vec \epsilon_d$ is the deuteron polarization vector.
At threshold, we can further simplify Eq.~(\ref{eq:dcs}) using:
\begin{eqnarray}
 \sqrt{s}= m_d + \mpi,
  \qquad
   |\vec p_{i}| = \sqrt{ \frac{s}{4}  - \mn^2 }\,,
\end{eqnarray}
where $m_N, m_d$ and $m_{\pi}$ stand for  the nucleon, deuteron and pion masses, respectively.
To include the effects of $NN$ interaction in the initial and final state
it is convenient to perform a partial wave decomposition of the amplitude
Eq.~(\ref{eq:avgampl}).

The  total near threshold  cross section of the $pp \to d \pi^+$ reaction in the
center-of-mass system  is conveniently parametrized as
\begin{eqnarray}
  \sigma = \alpha \eta + \beta \eta^3,
\end{eqnarray}
where $\eta$ is the outgoing pion momentum in the units of
the pion mass, i.e., $|\vec q_\pi|  = \eta \, m_\pi$.
The first term gives the outgoing s-wave pion contribution while the
second one corresponds to an outgoing p-wave pion.
At threshold, we only consider s-wave pion production, which means
the cross section becomes $\sigma = \alpha \eta$,
and the only contribution to $\alpha$ comes from the initial $pp$
$^3P_1$ partial wave in spectroscopic notation. 
We denote the threshold transition amplitude\footnote{
The  amplitude $M_{3P1}$  is related  to the  amplitude $\cal B$  used in
Refs.~\cite{Filin:2012za,Filin:2013uma} as  $M_{3P1}= 4i {\cal B}/\sqrt{3}$.
} as
$M_{3P1}$
\begin{eqnarray}\nonumber
M_{pp\to d\pi^+}&=&-\sqrt{\frac{3}{2}} M_{3P1}{\cal }\,(\vec{\mathcal{S}}\times \hat p_i\,)\cdot \vec\epsilon_d{\,^*}\ ,
\label{eq:ampproj}
\end{eqnarray}
where the relevant spin-angular structure of the initial and final nucleon pairs
are shown explicitly.
 Here 
$\vec {\mathcal{S}}=\chi^T_2{\sigma_y}\vec \sigma\chi_1/\sqrt{2}$
denotes the normalized spin structure of the initial spin-triplet    state and $\hat p_i = \vec p_i/p_i$.
We find
\begin{eqnarray}
\label{eq:alphaViaAmplitude}
\alpha =  \frac{4 \pi}{64 \pi^2} \frac{\mpi}{p_{i} (\mpi + m_d)^2} \frac{3}{4} |M_{3P1}|^2.
\end{eqnarray}

At threshold, the value of the parameter $\alpha$  can be extracted from the precise
pionic deuterium lifetime experiment  performed at PSI%
~\cite{Strauch:2010rm,Strauch:2010vu}, which has determined
the total cross section for the channel $nn \to d \pi^{-}$
to be $\sigma(nn \to d \pi^{-}) = ( 252^{+5}_{-11}) \eta \, \mu b$.
Using this value and neglecting isospin breaking effects,
we can extract the absolute value of the amplitude $|M_{3P1}|$ at threshold:
\begin{eqnarray}
  \label{eq:expabsampl}
  |M_{3P1}|^\text{exp.} = 21.5^{+0.2}_{-0.5}.
\end{eqnarray}
The value of this amplitude will be compared with our chiral-EFT-based prediction
to be presented below.

%%%%%%%%%%%%%%%%%%%%%%%%%%%%%%%%%%%%%%%%%%%%%%%%%%%%
\section{Formalism}
\label{sec:formalismandpc}
%%%%%%%%%%%%%%%%%%%%%%%%%%%%%%%%%%%%%%%%%%%%%%%%%%%%
 
The transition amplitudes for the $NN \to NN \pi$ processes involve several ingredients,
namely
the pion production operators from the two-nucleon system
as well as $NN$ bound and scattering states.
Since the $NN$ interaction is non-perturbative, these states cannot be directly calculated in
chiral perturbation theory. Here and in what follows, we employ
the {hybrid approach} suggested in~\cite{Weinberg:1992yk}.
The full amplitude is then calculated in two steps. First, the {irreducible pion production operator} is calculated
perturbatively in the chiral EFT framework.
In the second step, the resulting operator is {convoluted with $NN$ wave functions} obtained from
the solution of a non-perturbative Lippmann-Schwinger or Schr{\"o}dinger equation
with a realistic $NN$ potential. We will also show results based on
$NN$ potentials derived in the framework of chiral EFT. 
It is important to ensure that the production operator itself does not contain any parts of
$NN$ interaction in order to avoid double counting.
For this reason, only irreducible contributions to the production operator are considered,
i.e.~those which do not contain intermediate two-nucleon cuts.
Notice that  special care is required in order to isolate irreducible
contributions if the production operator
involves energy-dependent vertices,
see Ref.~\cite{Lensky:2005jc} for more details.

The pion production operator is computed perturbatively in the
framework of chiral EFT.\@ It was shown
in~\cite{Park:1995ku,Sato:1997ps,Hanhart:1997jd,Dmitrasinovic:1999cu,
Ando:2000ema,Bernard:1998sz,Kim:2007eh,Kim:2008qha}
that a naive application of the standard power counting rules, which
treat all typical momenta $p$ involved in a reaction on the same
footing as the pion mass $m_\pi$, fails to reproduce the data and does
not result in a convergent chiral expansion for the considered
process. It was suggested in Refs.~\cite{Cohen:1995cc,daRocha:1999dm}
that the power counting should be modified
to explicitly take into account the soft scale associated with typical
transferred momenta $p \sim  \sqrt{m_\pi m_N}$ in pion-production
reactions. In the modified power counting
which we refer to as the momentum counting scheme (MCS),
the expansion parameter is
\begin{equation}
\chi_{\rm MCS}\simeq\frac{p}{\Lambda_\chi} \simeq
\sqrt{\frac{\mpi}{m_N}} \simeq \frac{\mpi}{p} \, ,
\label{expansionpapar}
\end{equation}
where $\Lambda_\chi \sim m_\rho \sim m_N$ (with $m_\rho$ being the $\rho$ meson mass) refers to the expected breakdown scale of the
resulting chiral EFT approach.

Given the relatively high nucleon momenta in the initial state in
the pion production reaction, it may be advantageous to include
$\Delta$(1232) as an explicit degree of freedom, see Refs.~\cite{daRocha:1999dm,NNpiMenu}
for the first steps in this direction and  Refs.~\cite{Niskanen:1978vm,Hanhart:1998za} for
earlier phenomenological calculations. Since the
$\Delta$-nucleon mass splitting $\delta = m_\Delta - m_N$ does not vanish
in the chiral limit $m_\pi \to 0$, the resulting EFT has to be
regarded as a phenomenological extension of chiral perturbation
theory. A systematic treatment of the $\Delta$ resonance can be
carried out using different counting rules, see e.g.~Refs.~\cite{Hemmert:1997ye,Pascalutsa:2002pi}.  This
is expected to result in more natural values of the low-energy constants (LECs) in
the effective Lagrangian and an improved convergence of the EFT
expansion. Here and in what follows, we utilize the approach of
Ref.~\cite{Hanhart:2002bu} and count the $\Delta$-nucleon mass splitting
within the MCS as $\delta \sim p \sim \sqrt{m_\pi m_N}$.
Further, to systematically
investigate the role of the $\Delta$ isobar in the considered process, we
will implement different calculational strategies which are briefly outlined
below.
\begin{itemize}
\item
The most complete treatment of the $\Delta$ isobar is achieved by
using the $\Delta$-full formulation of chiral EFT, which is based on
the effective Lagrangian for pions, nucleons and  $\Delta$ isobars,
and by including  $\Delta$-excitations in the Hilbert space which results
in a coupled-channel framework. The initial state interaction
then includes the $NN\to N \Delta$ transition and, consequently,
the pion-production operator involves a contribution from the
$N \Delta \to NN \pi$ channel\footnote{
The intermediate  $\Delta\Delta$  state is  suppressed in our power counting
\cite{Filin:2013uma} and therefore not included explicitly in the coupled-channel formalism.
}.
A coupled-channel extension of the $NN$ amplitudes  to account for the $NN\to N \Delta$
transitions was developed in Ref.~\cite{Deltuva:2003wm}
on the basis of the  CD Bonn potential.
Another model, the  CCF  model,  uses the
coupled-channel folded diagrams formalism to include
$NN\to NN$ and $NN\to N\Delta$ transitions, see Ref.~\cite{CCF}.
The resulting framework will be referred to as
the coupled-channel $\Delta$-full chiral EFT approach (CC$\chi$EFT$-\Delta$).
\item
The coupled-channel formulation described above does explicitly take
into account the momentum scale $\sim \sqrt{(m_\Delta - m_N) m_N}$
corresponding to real $\Delta$ excitations in $NN$ collisions.
This (numerically) rather high momentum scale can be
integrated out. The coupled-channel treatment can then be avoided by perturbatively including all effects
associated with the $\Delta$-nucleon mass difference $\delta$ in the pion production operator and
requiring that the Hilbert space consists of only nucleonic states.
The resulting formulation will be referred
to as the $\Delta$-full chiral EFT approach ($\chi$EFT$-\Delta$).
Notice that the contributions to the scattering amplitude which
distinguish between the CC$\chi$EFT$-\Delta$ and $\chi$EFT$-\Delta$
approaches were argued
in Ref.~\cite{Filin:2013uma}
to be
of a higher order  than the one considered in our  calculation.
Since most of the modern phenomenological and chiral $NN$ potentials do not include
the $NN\to N \Delta$ transition explicitly,
we will use this strategy  to study  the influence of  different
$NN$ wave functions on the results. (We also will give results based
on the coupled-channel extension of the CD Bonn potential of Ref.~\cite{Deltuva:2003wm}.)
\item
Meanwhile, one may integrate out the $\Delta$ degrees of freedom already
at the level of the effective Lagrangian which leads to the standard
$\Delta$-less formulation of chiral EFT to be referred as $\chi$EFT$-
\Delta \palka$.  In this formulation, all effects of the
$\Delta$-resonance are  implicitly  taken into account via the
LECs of the effective Lagrangian.
\end{itemize}
It is important to keep in mind that (i) the LECs have a different
meaning and take different values in the $\Delta$-full
and $\Delta$-less formulations of chiral EFT and (ii) the
contributions to the irreducible pion production operator are
different in all three approaches. More precisely, the production
operator involves only
diagrams with nucleon lines in the $\chi$EFT$-
\Delta \palka$ approach, while diagrams involving $\Delta$ isobars
may yield different contributions  in the
CC$\chi$EFT$-\Delta$ and $\chi$EFT$-\Delta$ formulations due to the
different meaning of irreducibility. This issue will be discussed in detail
in the next section.

Finally, the convolution of the production operator with the initial and
final $NN$ wave functions
is done by first performing a partial wave projection of the pion-production operator.
The resultant operator is sandwiched between the $NN$ wave functions and
integrated up to a momentum cutoff of order 600--1000 MeV which
is of the order of the expected breakdown scale of the chiral EFT.\@
The equations defining the convolution procedure are given in
Appendix A of Ref.~\cite{Filin:2013uma}.

%%%%%%%%%%%%%%%%%%%%%%%%%%%%%%%%%%%%%%%%%%%%%%%%%%%
\section{The S-wave pion-production operator in $\chi$EFT$-\Delta$}
\label{sec:operators}
%%%%%%%%%%%%%%%%%%%%%%%%%%%%%%%%%%%%%%%%%%%%%%%%%%%

\begin{figure}[th]
\centering
\includegraphics[width=11.9cm]{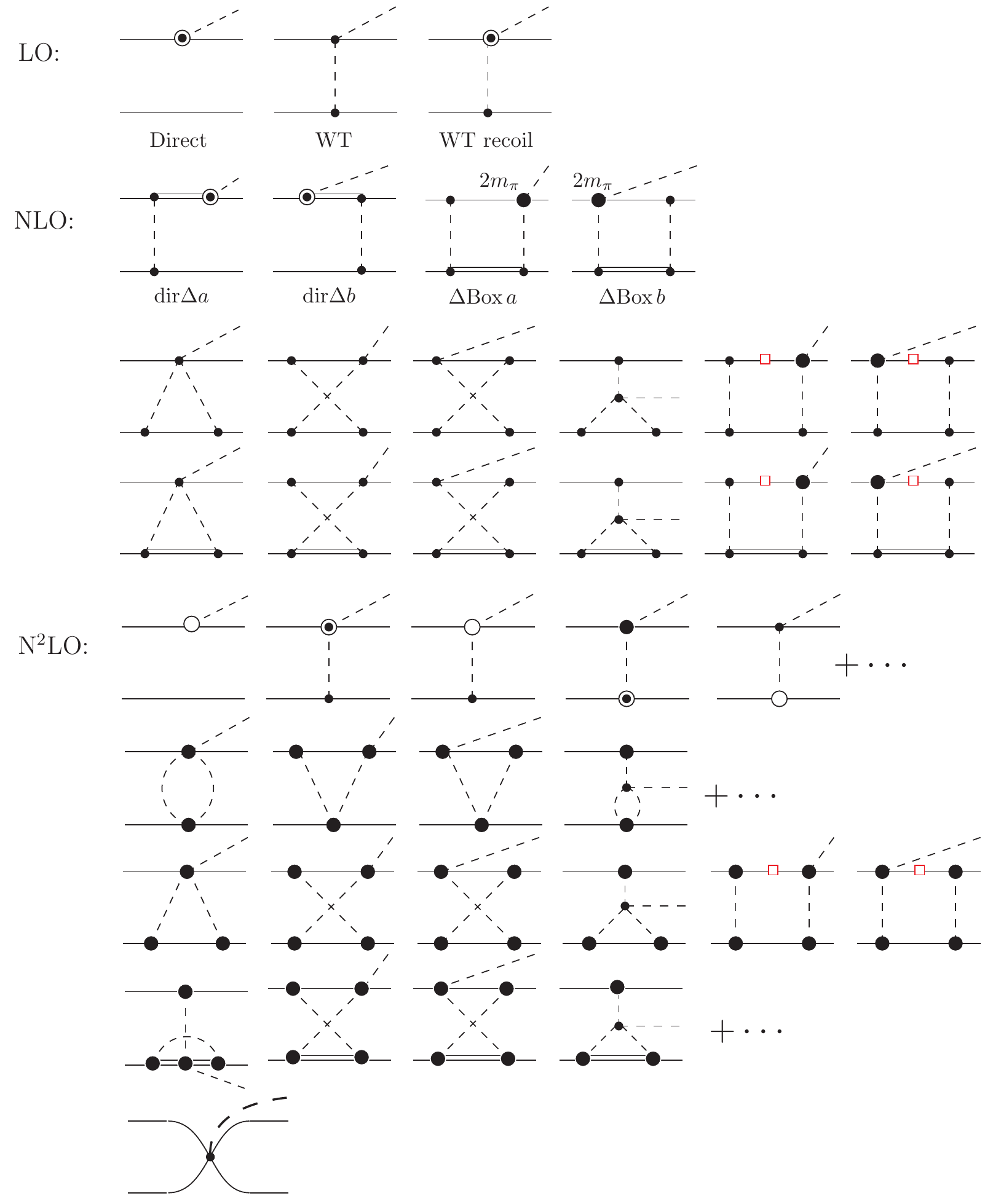}
\caption{\label{fig:allN2LO}
Diagrams contributing to the s-wave pion-production operator up to
NNLO in $\chi$EFT$-\Delta$.\@ 
Dashed, solid and double lines denote pions, nucleons and $\Delta$-resonance, respectively.
The complete expressions for the vertices and the corresponding Lagrangians are given in
Ref.~\protect\cite{Filin:2013uma}.
Solid dots refer to vertices from the leading-order Lagrangians ${\cal L}_{\pi N}^{(1)}$
and ${\cal L}_{\pi \pi}^{(2)}$ while vertices  denoted by the symbol $\odot$  originate from the
sub-leading Lagrangian ${\cal L}_{\pi N}^{(2)}$.
Filled circles  indicate the possibility to have
both leading and sub-leading vertices from  ${\cal L}^{(1)}_{\pi \! N}$ and
${\cal L}^{(2)}_{\pi \! N}$ in the diagram, see Fig.~2 in Ref.~\protect\cite{Filin:2013uma}
for clarification.
Open circles refer to
vertices from  ${\cal L}^{(3)}_{\pi \! N}$ while red squares on the nucleon propagator in the box diagrams indicate
that the corresponding nucleon propagator cancels with
parts of the $\pi N$ vertex and leads to the irreducible contribution,
see Ref.~\cite{Filin:2013uma} for further details.
The last NNLO diagram is the five-point contact term (CT) diagram.
}
\end{figure}

We have calculated the expressions for the pion production operator relevant for
s-wave pion production up-to-and-including NNLO in MCS within the
$\chi$EFT$-\Delta$ formulation.\@
The complete expressions for the operators were derived in Ref.~\cite{Filin:2013uma}
and are given in the Appendix~\ref{app:operators} of this  work,  while
in this section we summarize the most important results.
Notice further that the expressions for the production operator
in the $\chi$EFT$-\Delta \palka$ formulation can be obtained by simply
dropping the $\Delta$ contributions provided we redefine some of the LECs.

Only two types of diagrams contribute to
the dominant LO operators,
the single-nucleon ``direct'' pion production operator 
and the Weinberg-Tomozawa operator (including its recoil correction labelled WT recoil),
illustrated in the top row in Fig.~\ref{fig:allN2LO}.
At next-to-leading order (NLO), the  tree-level diagrams with intermediate
$\Delta$ excitations start to contribute
(first two diagrams  in the second row in Fig.~\ref{fig:allN2LO} labelled
as  ``dir$\Delta a$'' and ``dir$\Delta b$'').
In addition, loop diagrams appear at NLO.\@   However, as was shown in Refs.~\cite{Hanhart:2002bu,Lensky:2005jc}, the sum of
all NLO loop diagrams  illustrated in the third row
in Fig.~\ref{fig:allN2LO} cancel exactly for s-wave pion production.
Refs.~\cite{Hanhart:2002bu,Filin:2013uma} showed that also the NLO loop diagrams
with an intermediate
$\Delta$ shown in the fourth row  in Fig.~\ref{fig:allN2LO} cancel exactly.
Meanwhile,  there are also non-vanishing contributions of the  loop diagrams
shown  in the second row  in Fig.~\ref{fig:allN2LO}
(diagrams ``$\Delta $Box $a$'' and ``$\Delta $Box $b$'').
The contributions of  these box  diagrams  do not vanish
if the $\pi N\to \pi N $  vertex is  taken on shell,
that is  it should be proportional to $2m_{\pi}$
in full analogy to the    Weinberg-Tomozawa operator at LO.\@
Naively,   these operators  appear to be  suppressed  according to the MCS
and formally start to contribute at NNLO.\@
On the other hand,  these two ``box'' diagrams are exceptional among
loop operators in the sense that  their contributions  are potentially enhanced
due to the  presence of a   (reducible) $N\Delta$  intermediate state.
 Indeed,      due to the relatively small mass difference
between the nucleon and $\Delta$,
the $N\Delta$ propagator at the pion production threshold   effectively
scales  as
\begin{equation}
\frac{1}{m_{\pi}-\delta  - p^2/m_N} \sim
  \frac{1}{m_{\pi}},
\end{equation}
in contrast to a  $1/\delta\sim 1/p$ behavior expected from an  MCS estimate.
This  might lead to an  enhancement
compared to the expected MCS contribution
of these  box diagrams.
This argument is supported by the explicit calculations presented  in  Sec.\ref{sec:delta}.
Following this logic,  we promote  these two  particular box  diagrams
to NLO, i.e.~to  the order where there are   other  (tree-level)  diagrams
with  the  $N\Delta$  intermediate state
shown in the second line in Fig.~\ref{fig:allN2LO}\footnote{
Note that only the diagrams in the second row in Fig.~\ref{fig:allN2LO},
involving   an ``initial''  $\Delta$,  i.e.~dir$\Delta a$ and $\Delta $Box\,$a$,
contribute to $pp\to d\pi^+$  while all four   
diagrams are relevant for $pp\to pp\pi^0$.
}.
This treatment  of these diagrams is consistent with that
used in~\cite{Baru:2007wf,Baru:2011bw,Baru:2010xn}
 to calculate the  corrections to the pion deuteron scattering length  due to the $\Delta(1232)$.
 In particular,  it  was shown in Ref.~\cite{Baru:2007wf}
that both contributions  from the ``direct''  pion emission via
the  intermediate  $\Delta(1232)$ excitation
(the diagrams similar to the first two in the second line in Fig.~\ref{fig:allN2LO})
and  the $\Delta$-box diagrams involving  the Weinberg-Tomozawa
on shell $\pi N \to \pi N$ vertex
are roughly  of  similar size.
 
Finally, at NNLO there are tree-level and loop diagrams as well as the
five-point contact terms (CTs)
contributing to the pion production operator. Specifically,   there is   one contact term in  the  reaction
channel $pp\to d\pi^+$ and another  in $pp\to pp\pi^0$.
In addition, there are numerous NNLO loop diagrams with explicit $\Delta$'s.
Only some of these loop diagrams are  illustrated
in Fig.~\ref{fig:allN2LO}.
We have calculated all diagrams and found numerous cancellations among
the loop contributions at NNLO~\cite{Filin:2013uma}.\@
Unlike NLO loop diagrams (see rows 3 and 4 in Fig.~\ref{fig:allN2LO}), where the cancellation is exact, a finite contribution remains after
renormalization of the operators at NNLO.\@
We also find a finite contribution from the NNLO tree-level diagrams.
The complete set of analytic expressions for the NNLO pion production operators is
given in Ref.~\cite{Filin:2013uma} and is summarized in  appendix~\ref{app:operators}.

In the next section we consider the contributions of these operators to the
$pp \to d \pi^{+}$ threshold amplitude and compare them with the experimentally
determined amplitude in Eq.~(\ref{eq:expabsampl}).
As should be clear from the discussion in this section, our calculation is parameter free up-to-and-including NLO,
while at NNLO there is one contact  term which can always be  adjusted
to compensate the deviation from  the experimental amplitude at threshold.
The goal  of the study,  however,
is  to demonstrate that the counting  scheme used to  classify the operators
is adequate, i.e.~the size of the operators which appear  at  the given order is  in agreement
with the estimate.   This is a precondition for a reliable estimate of the theoretical uncertainty  and is
also needed to correctly identify the production mechanism  in the   chirally suppressed $pp\to pp\pi^0$ channel.
In order to comply with this goal,  in the next section we  discuss the  contribution of  the NNLO
operators  without the NNLO contact term  and  compare this result with  the  estimate expected
based on Eq.~\eqref{expansionpapar}.  Further,  from the  difference between the  NNLO  theoretical
prediction (without the contact term)  and  the data, we extract the  value of the contact term contribution
and  again confront it with the estimate.

%%%%%%%%%%%%%%%%%%%%%%%%%%%%%%%%%%%%%%%%%%%%%%%%%%%
\section{Numerical results}
\label{sec:convphenom}
%%%%%%%%%%%%%%%%%%%%%%%%%%%%%%%%%%%%%%%%%%%%%%%%%%%

We calculate the threshold amplitude $M_{3P1}$ for the reaction $pp \to d\pi^+$
 by performing the convolution of pion production operators  of section~\ref{sec:operators}
 (see Fig.~\ref{fig:allN2LO} and  also the appendix~\ref{app:operators})
 with a set of $NN$ wave functions
 derived from the modern phenomenological potentials: CD Bonn~\cite{Machleidt:2000ge},
 Nijmegen~\cite{Stoks:1994wp} and AV18~\cite{Wiringa:1994wb}.
In what follows, the expression for the $NN \to NN \pi$ operator, which is
derived based on the diagrams shown in Fig.~\ref{fig:allN2LO}, will
be called the (s-wave)
\emph{pion production operator}
while its convolution with the $NN$ interaction in the initial and final state
will be referred to as the (s-wave) \emph{pion production amplitude}.
In this section we discuss the results of the convolution and compare the
resulting amplitudes with the value extracted from experiment in Eq.~(\ref{eq:expabsampl}).
In our momentum space evaluations of the production amplitude,
the convolution integrals are supplied with a sharp  ultraviolet cutoff $\Lambda$, which
will be varied in a certain range.

\begin{figure}[t]
  \centering
  \begin{minipage}[b]{0.45\textwidth}
    \includegraphics[width=\textwidth]{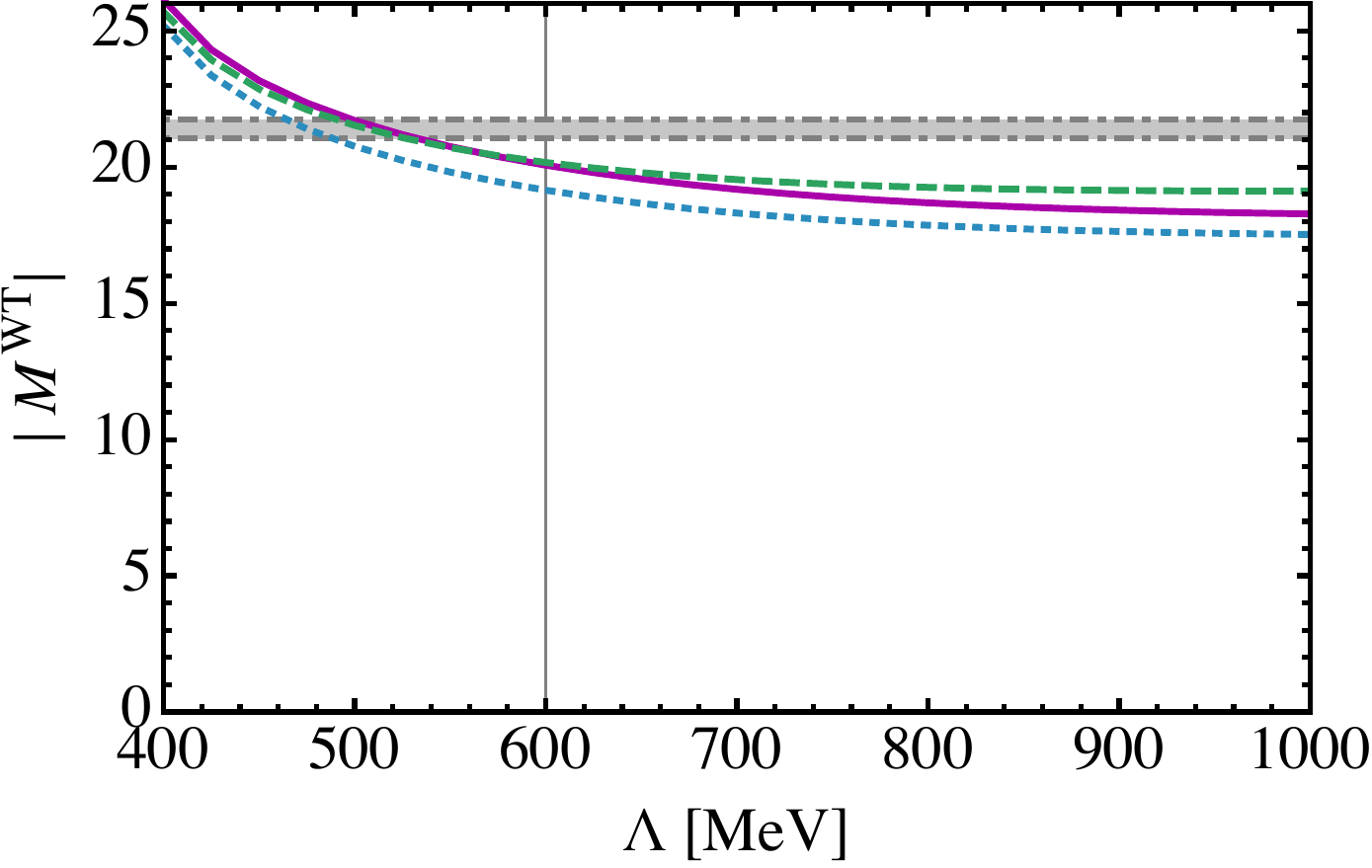}
  \end{minipage}
   \qquad
  \begin{minipage}[b]{0.45\textwidth}
    \includegraphics[width=\textwidth]{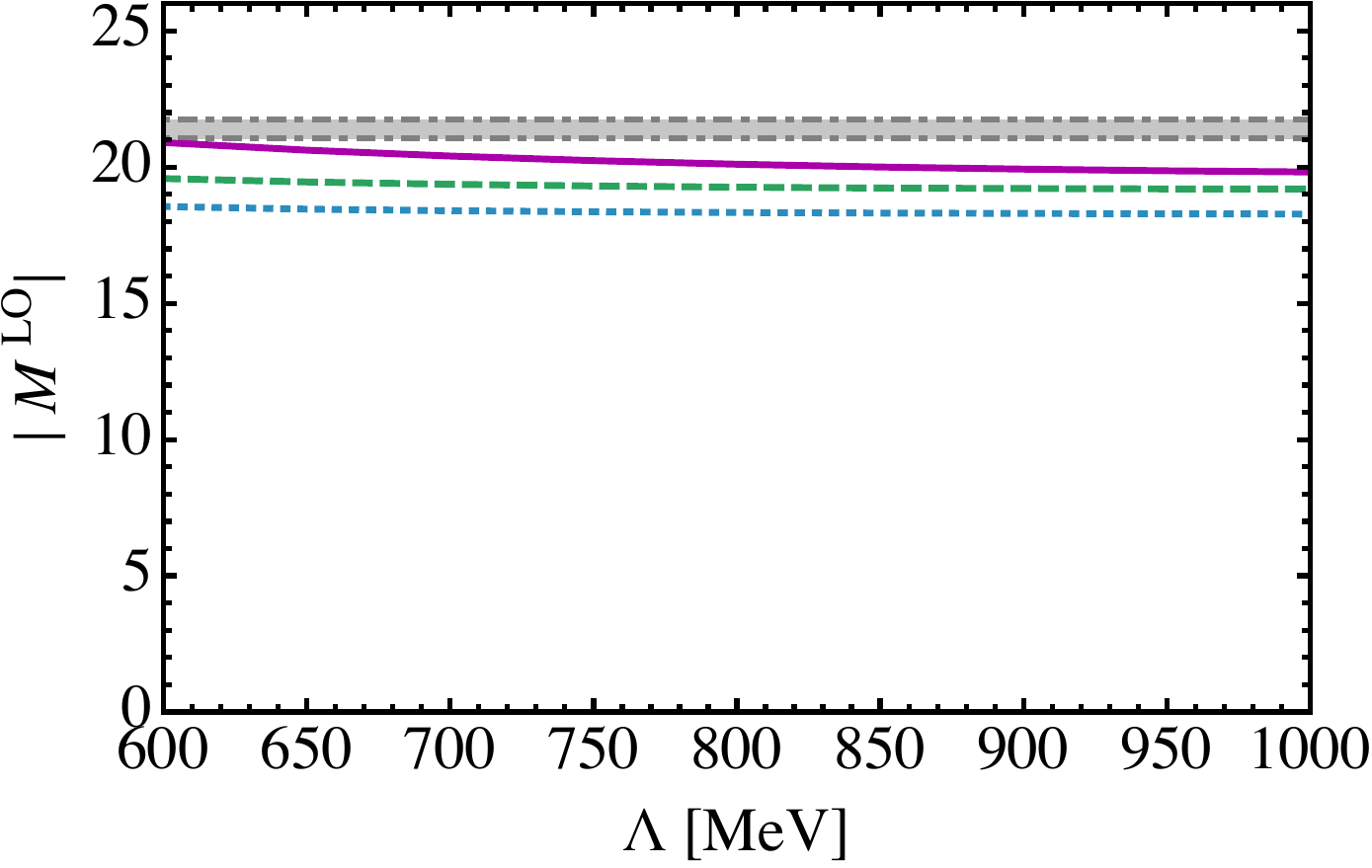}
  \end{minipage}
  \caption{\label{fig:amplWTandLO}
    The $pp \to d\pi^+$ amplitude $|M_{3P1}|$ calculated
    based on solely the LO
    Weinberg-Tomozawa operators (left panel)
    and all LO operators (right panel)
    as a function of a sharp momentum integral cutoff $\Lambda$.
    The amplitude is calculated using various \emph{phenomenological}
    $NN$ wave functions:  
    solid violet line --- CD Bonn~\cite{Machleidt:2000ge}, dashed green --- Nijm 1~\cite{Stoks:1994wp}, dotted blue --- AV18~\cite{Wiringa:1994wb}.
       The vertical line in the left panel indicates the value of the cutoff
    (about 600 MeV) where amplitude becomes (almost) cutoff independent.
      The horizontal  grey band  between the dot-dashed lines shows the experimental value of the amplitude
      $|M_{3P1}|$  including the errors (see Eq.~(\protect\ref{eq:expabsampl})) extracted from Refs.~\cite{Strauch:2010rm,Strauch:2010vu}.
  }
\end{figure}

To test the convergence of chiral EFT, we find it instructive to consider the
contributions of the various pion-production operators, given in the appendix~\ref{app:operators}, separately.
We start the discussion with the long-range leading-order pion-production
operator corresponding to the diagrams WT and WT  recoil  in
Fig.~\ref{fig:allN2LO}  involving the Weinberg-Tomozawa vertex.
This LO operator is known~\cite{Lensky:2005jc} to give the most important contribution
to the amplitude $M_{3P1}$.
In the left panel of Fig.~\ref{fig:amplWTandLO}, we show the contribution to
the amplitude from the WT diagrams
as a function of the cutoff $\Lambda$ calculated using various phenomenological $NN$ wave functions.
As expected for a long-range operator, the observable has almost no dependence
on the $NN$ interaction model for $\Lambda \ge 600$ MeV.
The cutoff-dependence for small cutoff values in the range 400--600 MeV is not surprising
since the typical momentum transfer in the pion production is about 360 MeV.
Therefore, for cutoffs less than about 600 MeV, the separation of
intermediate and short distance scales becomes
insufficient to get fully  cutoff-independent results.
In what follows we consider cutoff values  in the range   $\Lambda=$ 600--1000 MeV.

\begin{figure}[t]
  \centering
  \includegraphics[width=.45\textwidth]{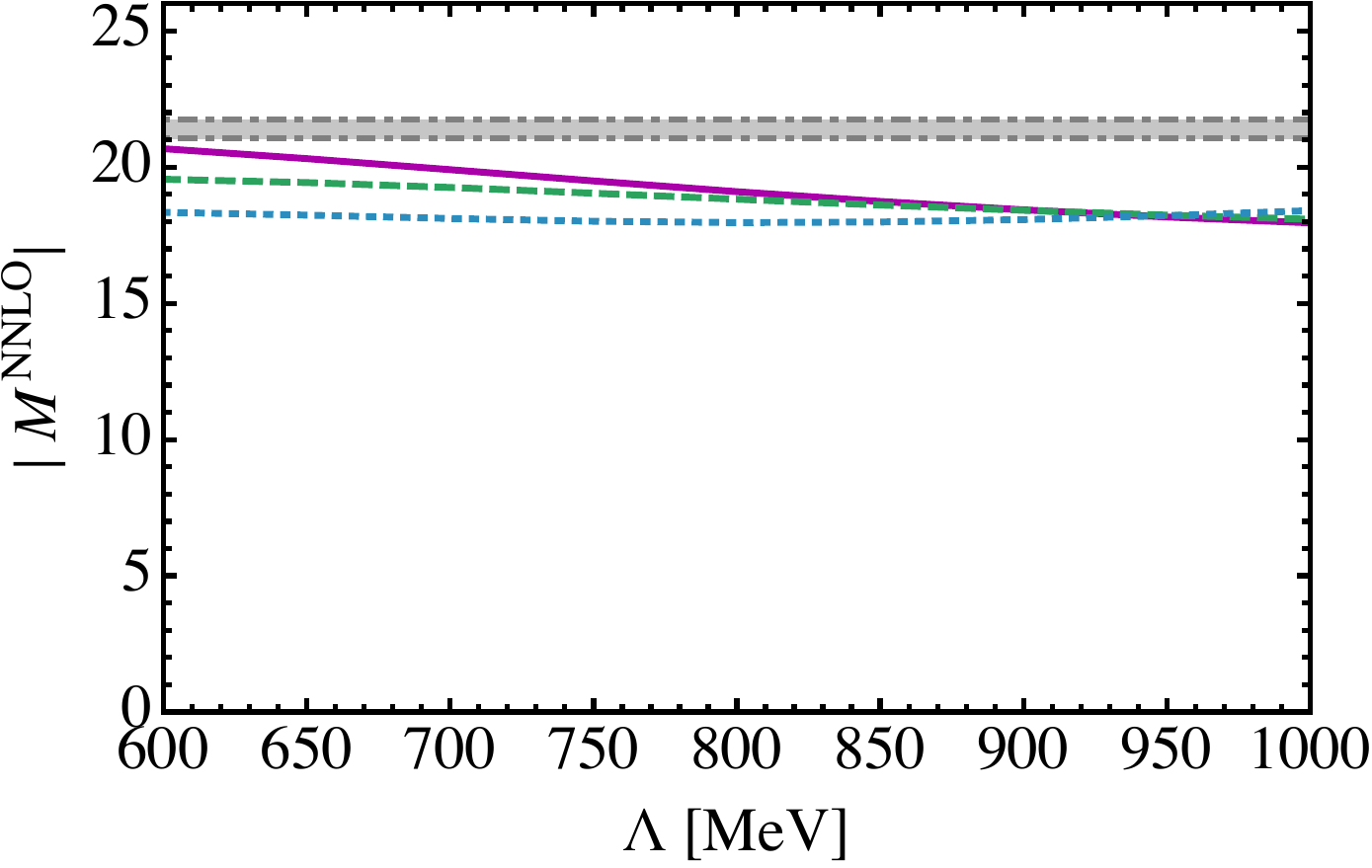}
  \caption{\label{fig:amplNNLO}
    The $pp \to d\pi^+$ amplitude $|M_{3P1}|$ as a function of the cutoff
    calculated using the complete set of LO, NLO and  NNLO operators
    \underline{except} for the NNLO five-point contact term operator.
    The amplitude is calculated using several \emph{phenomenological} $NN$ wave functions,  see Fig.~\ref{fig:amplWTandLO} for notation.
    The difference between experimental  and evaluated amplitudes
    determines the    required strength of the unknown five-point NNLO contact term
    in order for the theory to fit the data.
  }
\end{figure}
 
Next, we consider the complete leading-order result.
In addition to the WT operator (and its recoil correction), there is one more diagram at LO, \@
which is known as the direct pion production operator (Fig.~\ref{fig:allN2LO}).
Since this is a single-nucleon operator, which probes the $NN$ interaction at shorter distances,
it is natural to expect that the total LO amplitude might become
more sensitive to the short-range details of the different $NN$ models.
However, the contribution of the direct term is known to be very small due to
a destructive interference
between the Born term and the contribution of the $NN$ initial state
interaction, see e.g.~Ref.~\cite{Lensky:2005jc} where this interference was shown for
CCF~\cite{CCF} and CD Bonn~\cite{Machleidt:2000ge} potentials.
As a consequence, the direct pion production operator does not introduce any
apparent
cutoff dependence for the three $NN$ potentials
as can be seen in the right panel of the Fig.~\ref{fig:amplWTandLO}.
The evaluated complete LO amplitude is  close to but smaller than
the experimental value for $M_{3P1}$
in Eq.~(\ref{eq:expabsampl}) for all phenomenological $NN$ potentials used
in our calculations.

We are now in the position to consider the complete pion production amplitude including
all LO, NLO and NNLO pion production operators
introduced in section~\ref{sec:operators}.
First, we note that    all NLO operators from the loop diagrams were shown
to cancel exactly in Ref.~\cite{Lensky:2005jc}.
This cancellation also applies to the NLO loop diagrams involving the intermediate $\Delta$
as shown in Ref.~\cite{Filin:2013uma} (fourth row of diagrams in Fig.~\ref{fig:allN2LO}).
Further,   the  tree-level NLO operators from diagrams in the second row of
Fig.~\ref{fig:allN2LO}
show a destructive interference with the corresponding $\Delta$-box terms and their
net contribution is small,  see Sec.\ref{sec:delta}
for a more detailed discussion of  this cancellation.

Apart from the unknown LEC of
the five-point  CT (last  diagram in Fig.~\ref{fig:allN2LO}),
we need to specify the values of the LECs $c_i$, $i$ = 1, \ldots, 4, in our
evaluation of the production amplitude at NNLO.\@
The values for the $c_i$'s are taken from   a tree-level order fit  (chiral order $Q^2$ fit 1)
to $\pi$N scattering data, Ref.~\cite{Krebs:2007rh}.
The sensitivity to  the choice of  these LECs is   discussed in Sec.~\ref{c_i}.

In Fig.~\ref{fig:amplNNLO}  we show the pion production amplitude
up-to-and-including NNLO as a function of the cutoff.
Comparing the results in  Figs.~\ref{fig:amplWTandLO} (right) and~\ref{fig:amplNNLO},
we conclude that the NNLO amplitude contributes up to  10\%
to the amplitude $|M_{3P1}|$. This result shows that the perturbative treatment of the
production operator is reasonable and supports the counting scheme used to classify the
pion production operators.
We also observe from Fig.~\ref{fig:amplNNLO}
that the results for all phenomenological $NN$ models are
in a reasonably good agreement with each other --- the changes in the results due to
the use of  different $NN$ models and from  the cutoff variations  are  well within
the NNLO estimates based on the  MCS.\@
We emphasize, however, that the contribution of the NNLO contact term (CT)
[see last row in Fig.~\ref{fig:allN2LO}] is \underline{not} included.
As usual in EFT,  this contact term is expected to compensate for this
(already rather mild) cutoff dependence as well
as for the natural dependence of the results from different   $NN$-models.
The difference between the calculated pion production amplitude,
which contains all LO, NLO and NNLO operators,  and the
experimental value of $|M_{3P1}|$ (shown as a horizontal dashed-dotted band)
illustrates the magnitude of the CT amplitude  required to reproduce the data.
In chiral EFT, the CT operators parameterize the contributions from short-range processes,
which are not treated as explicit degrees of freedom. For example,
short-range contributions due to exchange of the vector-mesons $\rho$- and $\omega$
are implicitly accounted for via the corresponding
NNLO CT operator in our chiral EFT.\@
A fit to the data reveals that the contribution of the counter term is of the order of 5--15\% of the LO contribution
(depending on the NN model used) which is   consistent with the power counting.
Finally, from the comparison of the LO result in Fig.~\ref{fig:amplWTandLO} (right)
with the experimental data,
we conclude that the net contribution from all NNLO operators (including the contact term)
is very small and  fully in line with the power counting.
 
%%%%%%%%%%%%%%%%%%%%%%%%%%%%%%%%%%%%%%%%%%%%%%%%%%%
\subsection{Sensitivity to the LECs $c_i$}
\label{c_i}
%%%%%%%%%%%%%%%%%%%%%%%%%%%%%%%%%%%%%%%%%%%%%%%%%%%

 We now address the sensitivity of the pion production amplitude
to the values of the LECs $c_i$.
As already mentioned,  the results  shown in Fig.~\ref{fig:amplNNLO}
are obtained with
 the $c_i$ values determined from  a tree-level
 order fit  (chiral order $Q^2$ fit 1)
 to $\pi$N scattering data~\cite{Krebs:2007rh}  in a theory with an explicit $\Delta$,  namely
 \begin{equation}
c_1=-0.57\mbox{ GeV}^{-1},  \quad  c_2= -0.25\mbox{ GeV}^{-1}, \quad c_3 = -0.79\mbox{ GeV}^{-1}, \quad c_4= 1.33\mbox{ GeV}^{-1}.
\label{ciQ2}
\end{equation}
These values correspond to the $\pi N\Delta$ decay  constant  $\gpind=1.34$.
 
On the other hand,  the values of the $c_i$'s emerging from one-loop
calculations of $\pi N$ scattering are well-known to be significantly
different
\cite{Fettes:1998ud,Krebs:2012yv,Fettes:2000xg,Alarcon:2012kn,Chen:2012nx,Wendt:2014lja,Siemens:2016hdi}.
For a discussion  of the LECs  $c_i$ extracted from $NN$ analyses we refer to
Ref.~\cite{Epelbaum:2014efa}. Notice further that pion-nucleon
scattering phase shifts were recently determined in the framework of the
Roy-Steiner equation which takes into account constraints  from
analyticity, unitarity, and crossing symmetry~\cite{Hoferichter:2015dsa}. It is
conceivable that this analysis will allow for a more accurate
determination of the $c_i$'s in the future, see~\cite{Hoferichter:2015tha} for a first step along this
line.  To have an idea of the sensitivity of our results to the values
of these LECs, we consider the empirical values of
\begin{equation}
c_1=-0.81\mbox{ GeV}^{-1},  \quad  c_2= 3.28\mbox{ GeV}^{-1}, \quad c_3 = -4.69\mbox{ GeV}^{-1}, \quad c_4= 3.40\mbox{ GeV}^{-1},
\label{empiricFull}
\end{equation}
which have been used in the new generation of chiral $NN$ potentials
at NNLO and N$^3$LO of Ref.~\cite{Epelbaum:2014efa}. Except
for $c_2$, these values correspond to the analysis of $\pi N$
scattering inside the Mandelstam triangle of Ref.~\cite{Buettiker:1999ap}, where the chiral expansion
is expected to converge faster than in the physical region.
The value of $c_2$, which could not be determined reliably in
Ref.~\cite{Buettiker:1999ap},
is taken from    the one-loop $Q^3$ calculation of  Ref.~\cite{Fettes:1998ud}.
Given that the  values of the $c_i$'s in Eq.\eqref{empiricFull} have been obtained in the
standard formulation of chiral perturbation theory based on pions and
nucleons as the only explicit degrees of freedom, we have to subtract
the leading  $\Delta$ contributions in order to be able to use the
LECs in $\chi$EFT$-\Delta$. Using the well-known $\Delta$ contributions
discussed in Ref.~\cite{Bernard:1996gq}, we arrive at the values of
\begin{equation}
c_1=-0.81\mbox{ GeV}^{-1},  \quad  c_2= 0.56\mbox{ GeV}^{-1}, \quad c_3 = -1.97\mbox{ GeV}^{-1}, \quad c_4= 2.04\mbox{ GeV}^{-1}.
\label{empiric}
\end{equation}
These numbers are used in our calculation together with   $\gpind=1.34$.

\begin{figure}[t]
  \centering
  \includegraphics[width=.45\textwidth]{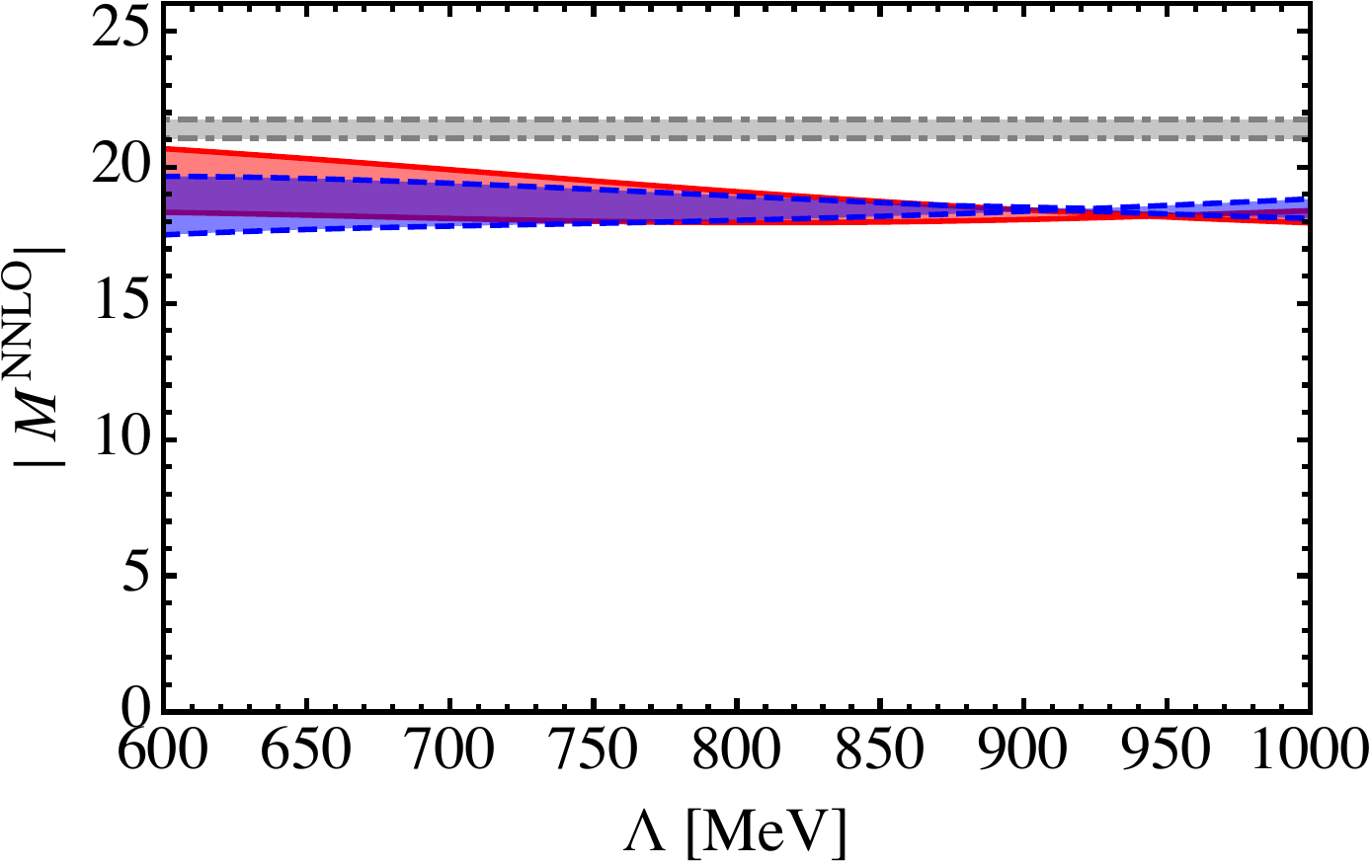}
  \caption{\label{fig:amplNNLO-ciValues}
Sensitivity of the  $pp \to d\pi^+$ amplitude to the choice of the LECs $c_i$.
    The red band restricted by  two solid lines  corresponds to the results obtained with
    the order-$Q^2$ values of the $c_i$'s specified in Eq.~(\ref{ciQ2})
    for three phenomenological $NN$ potentials (AV18, Nijm1 and CD
    Bonn).
    The blue  band between the dashed lines is obtained using  the
    empirical  values of the $c_i$'s~\cite{Buettiker:1999ap,Epelbaum:2014efa} with the $\Delta$
    contributions being subtracted as specified in Eq.~(\ref{empiric}) and
    employing the same $NN$ wave functions.
   The NNLO contact term is \underline{not} included.
        See Sec.~\ref{c_i} for a more complete discussion and Fig.~\ref{fig:amplWTandLO} for notation  of
    the horizontal  grey band.
  }
\end{figure}

In Fig.\ref{fig:amplNNLO-ciValues} we  illustrate the effect of   the variations of the LECs $c_i$
on  the  $pp \to d\pi^+$ amplitude.
For this purpose,  we  calculate the  pion-production amplitude as a function of the
 cutoff with the two sets  of the  $c_i$'s  as specified in Eqs.~(\ref{ciQ2}) and (\ref{empiric})  for
 several  $NN$ potentials.
 Specifically,  the red band  is restricted by  two solid lines obtained    using  the
 three different $NN$ models corresponding to  the AV18, Nijm1 and
 CD Bonn potentials~\cite{Machleidt:2000ge,Stoks:1994wp,Wiringa:1994wb},
 with  the order  $Q^2$ values of the $c_i$'s  from Eq.~(\ref{ciQ2}).
Similarly,  the blue  band is  calculated  with the same $NN$ potentials  but
with the empirical  values of the $c_i$'s  from Eq.~(\ref{empiric}).
We  conclude that  the  difference  between the bands
lies well within the uncertainty estimate expected at  order NNLO.\@

%%%%%%%%%%%%%%%%%%%%%%%%%%%%%%%%%%%%%%%%%%%%%%%%%%%
\subsection{Effects of  the  $\Delta(1232)$ on the  threshold pion-production amplitude}
\label{sec:delta}
%%%%%%%%%%%%%%%%%%%%%%%%%%%%%%%%%%%%%%%%%%%%%%%%%%%

As  discussed in Sec.~\ref{sec:formalismandpc},
the inclusion of the $\Delta$  degree of  freedom in the theory may be accomplished using
the two different strategies described in Sec.~\ref{sec:formalismandpc}.
The results already presented correspond to the $\chi$EFT$-\Delta$
formulation, where the contributions of the
$\Delta$  resonance are perturbatively
incorporated into the pion-production operator as  shown
in the second row in Fig.~\ref{fig:allN2LO}.  
This operator is then sandwiched by  the  initial and final $NN$  state wave  functions.
Note that the $NN-N\Delta$ transition, which naturally appears as  a part of the pion-production operator,
contains a  contact term in addition to  the  one-pion exchange (OPE) potential.
However, this contact term appears to be suppressed by $\chi_{\rm MCS}^2$ (see Eq.~\eqref{expansionpapar}) relative to  
the OPE potential, since it comes with at least two derivatives as a consequence of the Pauli principle
(see, e.g., Ref.~\cite{Epelbaum:2007sq} for a related discussion).
Since the  $\Delta$  in  $NN\to NN\pi$ starts to contribute  at NLO,  to the order
we are  working, the  unknown short  range part in the $NN\to N\Delta$ transition    can be  dropped.
 Hence,  only  the  diagrams  shown in   the second row in Fig.~\ref{fig:allN2LO} are relevant.

\begin{figure}[t]
\includegraphics[scale=0.6]{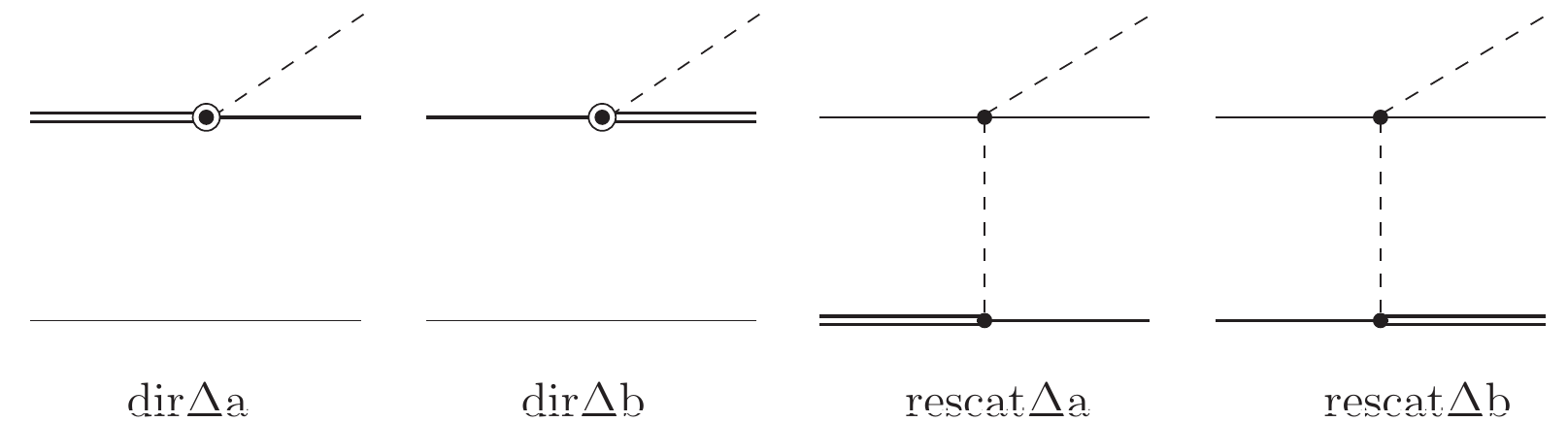}
\caption{\label{fig:treeDelta}
Additional diagrams contributing to the production operator in the
CC$\chi$EFT$-\Delta$ formulation. In the last  two rescattering
diagrams, only  the  on-shell  part  of   the  $\pi N$  scattering vertex
(2\mpi)  should  be  included~\cite{Filin:2013uma}.
Note that only  the diagrams,  which contain  $\Delta$ in the initial state,  i.e.,
dir$\Delta$a  and rescat$\Delta$a,  contribute to the reaction $pp\to d\pi^+$.}
\end{figure}

In the CC$\chi$EFT$-\Delta$ formulation, the initial
$NN$ and $N\Delta$ states are generated non-perturbatively.
 The corresponding elastic $NN\to NN$ and inelastic $NN \to N \Delta$
  transition amplitudes are  obtained   as  a   solution
  of the  coupled-channel  system  with $NN$ and  $N\Delta$    interactions
   \begin{eqnarray}\nonumber
     T_{\rm NN}&=&  V_{\rm NN}+  V_{\rm NN} G_{\rm NN}  T_{\rm NN} +
     V_{\rm NN-N\Delta} G_{\rm N\Delta}  T_{\rm N\Delta-NN}, \\
     T_{\rm N\Delta-NN}&=&  V_{\rm N\Delta-NN}+  V_{\rm N\Delta-NN} G_{\rm NN}  T_{\rm NN} +
     V_{\rm N\Delta-N\Delta} G_{\rm N\Delta}  T_{\rm N\Delta-NN},
     \end{eqnarray}
where  $G_{\rm NN} (G_{\rm N\Delta})$  is the  $NN$  ($N\Delta$) Green-function
and
$V_{\rm NN}$,  $ V_{\rm N\Delta-N\Delta}$,  and $V_{\rm N\Delta-NN}$
are the corresponding  elastic  and transition potentials, respectively.
The short-range parts of  the   $N\Delta$ and $\Delta\Delta$ interactions  are
constrained   by  fitting  the $NN$   observables~\cite{Deltuva:2003wm,CCF}.
Since  the $NN$  and $N\Delta$  states are coupled,
the  full pion-production amplitude  also receives  contributions   from
diagrams  containing  initial and final $N\Delta$  states as    shown in
Fig.~\ref{fig:treeDelta}.
In a full  analogy  to the  ``direct''  single-nucleon diagrams in Fig.~\ref{fig:allN2LO},
diagrams shown  in  Fig.~\ref{fig:treeDelta}    do  not  contribute  to
the on-shell  pion-production operator  but  have to be taken into
account when convolved
with the  $NN-N\Delta$  wave functions  either  in the  initial
or in the  final state.
Notice that for  the $pp\to d \pi^+$ reaction, the  $NN-N\Delta$ transition can
only appear in the initial state due to isospin conservation.
To avoid  double counting,
all diagrams shown in the second row in Fig.~\ref{fig:allN2LO}  have
to be dropped when calculating the production operator in the
CC$\chi$EFT$-\Delta$ formulation.

We now compare the results of  both $\Delta$-full formulations with
each other and with the results using the $\Delta$-less approach.  For
the sake of a meaningful comparison, we use here the  wave functions
based on the CD Bonn potential~\cite{Machleidt:2000ge}
and on the coupled-channel version of the CD Bonn potential~\cite{Deltuva:2003wm}.
The resulting values of the  pion production amplitude $M_{3P1}$,
which do not include the contribution of the $NN\to NN\pi$ counter term,
are  collected in Table~\ref{Tab:Deltares}.  The calculations are done with the cutoff $\Lambda=1$ GeV.
 Notice that  the results with the explicit $\Delta$ degree of freedom
 ($\chi$EFT$-\Delta$ and CC$\chi$EFT$-\Delta$ in Table~\ref{Tab:Deltares})  are  obtained using
 the order-$Q^2$ values for the $c_i$'s  in Eq.~(\ref{ciQ2}), whereas  in the $\Delta$-less approach
 ($\chi$EFT$-\Delta \palka$)  we take the corresponding order-$Q^2$ values which include
 the contributions of the $\Delta$ isobar, namely~\cite{Krebs:2007rh}:
\begin{equation}
c_1=-0.57\mbox{ GeV}^{-1},  \quad  c_2= 2.84\mbox{ GeV}^{-1}, \quad c_3 =-3.87\mbox{ GeV}^{-1}, \quad c_4= 2.89\mbox{ GeV}^{-1}.
\label{ciQ2noD}
\end{equation}

We find that  the coupled-channel approach  (CC$\chi$EFT$-\Delta$) yields the
amplitude  which is  about $12\%$ larger
than the one in the $\chi$EFT$-\Delta$ formulation.
This difference is comparable with
an estimate of   the
NNLO  contributions  based on the  MCS.\@
This result indicates that the coupled-channel dynamics and  the inclusion of the short-range
$ N\Delta$ interaction, constrained by the $NN$ data,
is somewhat more important than
what is expected based on dimensional analysis.
 As  shown in Table~\ref{Tab:Dreduce},   this difference  can be  attributed to
 the contributions from the  diagrams   shown in the second row in
 Fig.~\ref{fig:allN2LO}  and  in  Fig.~\ref{fig:treeDelta},   all  involving the intermediate $N\Delta$ state.

  The contributions from the
 diagrams dir$\Delta$a   and $\Delta$Box\,$a$  in Fig.~\ref{fig:allN2LO} 
are  relatively  large in the $\chi$EFT$-\Delta$ framework,   as expected at NLO in the MCS, but they
interfere destructively.  The contributions of the
 direct and  rescattering  diagrams (dir$\Delta$a   and rescat$\Delta a$)  in the CC$\chi$EFT$-\Delta$ approach 
 are  smaller individually  than in the previous case, but  they
 also undergo significant  cancellations in the sum.
This finding was  also observed in Ref.~\cite{NNpiMenu} using the CCF model~\cite{CCF}.
The pattern shown in Table~\ref{Tab:Dreduce}  could have been  expected:   as is
known  from phenomenological studies,
the destructive interference  between  tensor parts of  the OPE  and short range potentials
leads to smaller amplitudes in the coupled-channel framework as compared  to a  perturbative treatment  where
no short-range term is included. As shown in Table~\ref{Tab:Dreduce}, the net contribution
of  the operators with the $N\Delta$ intermediate state  appears to be
small in both cases but  has  an opposite sign.
This sign difference accounts for the deviation between the results of two  $\Delta$-full approaches.
Interestingly,  the net contribution from  all other operators   in Fig.~\ref{fig:allN2LO},
apart those  with the $N\Delta$ intermediate state discussed above,  appears to be
almost insensitive to whether  $NN$  interaction in the initial state is treated using
 a full coupled-channel  approach (CC$\chi$EFT$-\Delta$) or    the Hilbert state  consists of only nucleonic states while
  $\Delta$  is included perturbatively ($\chi$EFT$-\Delta$).
Finally,  as seen in Table~\ref{Tab:Deltares},
if  the $\Delta$ degree of freedom is integrated out at the level of the effective Lagrangian ($\chi$EFT$-\Delta \palka$),
the  result for the  reaction amplitude at $\Lambda\simeq 1$~GeV is about  $10\%$  smaller  compared to
the CC$\chi$EFT$-\Delta$ formalism   but only  a few percent  
larger than the one in the $\chi$EFT$-\Delta$ framework.
It should be noted at this point that for the  cutoff $\Lambda\simeq 1$ GeV   the amplitudes discussed in Table~\ref{Tab:Deltares} are   already largely saturated.
On the other hand,  at smaller    cutoffs the NNLO amplitudes, which do not yet include the $NN\to NN\pi$ contact term contribution,  are  expected  to possess some 
cutoff dependence.  The individual numbers shown in Tables~\ref{Tab:Deltares} and~\ref{Tab:Dreduce} may therefore change.  
However, the destructive interference  pattern discussed above generally persists.  
Due to  this interference,  the difference  between  the $\Delta$-full  and     $\Delta$-less approaches, which is  estimated to be of  the size of NLO terms,  
is   smaller  for cutoffs  $\Lambda \ge 700$ MeV.  Moreover,     the  deviation in  the  
results for    all three  approaches  constitutes generally an NNLO  effect in this range of cutoffs.

It should be noted, however,  that the cancellation  discussed above is  probably a  particular  feature of   
s-wave  pion production in $pp\to d\pi^+$ channel.   For example,  the p-wave pion production 
amplitudes in this reaction channel   (especially the dominant  amplitude in the  $^1D_2\to {^3}S_1 p$ partial wave)
 do acquire a  large (NLO) contribution  from    diagram dir$\Delta$a    in Fig.~\ref{fig:allN2LO}   \cite{Baru:2009fm}
while    box diagram $\Delta$Box\,$a$   starts to contribute at  next-to-next-to-next-to-leading order 
(NNNLO)  only and  therefore is  expected to be suppressed.    It remains to be seen if the cancellation  discussed above 
takes place for s-wave pion production in   $pp\to pp\pi^0$  channel. 

\begin{table}[t]
\caption{ The results for the  amplitude $M_{3P1}$  calculated for three different  strategies with respect  to the treatment of the $\Delta$, as discussed in the text.  The results correspond 
to the cutoff $\Lambda = 1$~GeV. 
\label{Tab:Deltares}}
\begin{tabular*}{\textwidth}{@{\extracolsep{\fill}}c|c|c}
\hline
 CC$\chi$EFT$-\Delta$
                                                       &$\chi$EFT$-\Delta$  

  & $\chi$EFT$-\Delta \palka$   \\
\hline
$18.1 - 9.6 i$ &    $16.0 - 8.5 i$   &$16.5 - 8.8 i$\\
\hline
\end{tabular*}
\end{table}

\begin{table}[t]
\caption{Individual contributions to the  amplitude $M_{3P1}$  from the diagrams
with reducible $N\Delta$  intermediate states, as shown in
Figs.~\ref{fig:treeDelta}
(dir$\Delta a$ and rescat$\Delta a$) and~\ref{fig:allN2LO} (dir$\Delta a$ and $\Delta$Box\,$a$),
after the convolution with the appropriate initial state wave functions.
Diagrams ``b''  in the corresponding figures do not contribute to  the
reaction  channel with the isospin 0 (deuteron) final state.
Except  those diagrams discussed  above,
the net  contribution of all  other diagrams in Fig.~\ref{fig:allN2LO}
is labelled  as  ``All other diagrams''.  The results correspond 
to the cutoff $\Lambda = 1$ GeV. 
\label{Tab:Dreduce} }
\begin{tabular*}{\textwidth}{@{\extracolsep{\fill}}r|r}
\hline
%\hline
\noalign{\smallskip}  CC$\chi$EFT$-\Delta$      
 &     $\chi$EFT$-\Delta$    
 \\
   \hline
  dir$\Delta a$ = $2.8 - 1.5 i$ &  dir$\Delta a$ = $4.8 - 2.5 i$  \\
 rescat$\Delta a$ = $-1.9 + 1.0 i $ &    $\Delta$Box\,$a$ = $-6.1 + 3.2 i$\\
    All other diagrams= $17.3 - 9.2 i $ & All other diagrams= $17.3 - 9.2 i$\\
  \hline
\end{tabular*}
\end{table}

%%%%%%%%%%%%%%%%%%%%%%%%%%%%%%%%%%%%%%%%%%%%%%%%%
\section{Convolution with chiral $NN$ potentials}
\label{sec:convchiral}
%%%%%%%%%%%%%%%%%%%%%%%%%%%%%%%%%%%%%%%%%%%%%%%%%

In this section we  consider the convolution of pion-production
operators with $NN$ wave functions
generated by potentials derived in chiral EFT.\@ Certainly,
this approach is more consistent from the conceptual point of view than the
hybrid method employed in the previous sections since all ingredients
are calculated based on the same effective Lagrangian.
On the other hand, the available chiral nuclear potentials are
actually derived in the formulation of chiral EFT, where the momentum scale
$p \sim \sqrt{m_N m_\pi}$ associated with radiative pions and relevant for the pion production reaction is
integrated out.  Thus, it is, in fact, more appropriate to also regard such
calculations as of being a hybrid type in spite of the fact that the
corresponding $NN$ wave
functions are calculated in the framework of chiral EFT.\@
Furthermore, the soft nature of chiral EFT potentials corresponding to lower
values of the momentum-space cutoff $\Lambda_{NN}$ as compared with phenomenological
potentials suggests a possible appearance of significant
finite-$\Lambda_{NN}$ artefacts. We further emphasize that the initial
energy corresponding to the pion production threshold is at the very
edge of the applicability range of even the state-of-the-art
fifth-order chiral potentials of Ref.~\cite{Epelbaum:2014sza}.

Here and in what follows, we will use the new generation of $NN$
potentials up to fifth order (N$^4$LO) in the chiral expansion presented in
Refs.~\cite{Epelbaum:2014sza,Epelbaum:2014efa}. In contrast  to the
first-generation chiral N$^3$LO  NN forces of Refs.~\cite{Entem:2003ft,Epelbaum:2004fk}, the new
potentials utilize a coordinate-space regularization scheme for
long-range components which   reduces the amount of finite-regulator artefacts.
The employed coordinate-space cutoff is varied in the range $R = 0.8
\ldots 1.2$~fm,
which in momentum space roughly corresponds to cutoffs of the order of
$\Lambda_{NN}\sim 500 \ldots 330$~MeV. 
In contradistinction to the exponentially falling high-momentum behavior of the older chiral potentials,
the new chiral potentials used in this work   have a power-like momentum cut-off.
The new chiral potentials
preserve the correct analytic structure of the amplitude at low energies
and lead to a  good description of deuteron properties and $NN$ phase
shifts for the harder cutoff choices of  $R = 0.8
\ldots 1.0$~fm. For the two softest choices of the regulator with
$R=1.1$~fm and $R=1.2$~fm, one observes significant regulator
artefacts (especially at higher energies), see
Refs.~\cite{Epelbaum:2014sza,Epelbaum:2014efa} for more
details. Notice further that the corresponding values of the momentum-space
cutoff, $\Lambda_{NN} \sim 360$~MeV and  $\Lambda_{NN} \sim 330$~MeV,
are  comparable with or even smaller than the momentum transfer scale $ p
\sim \sqrt{m_N m_\pi} \sim 360$~MeV inherent in pion production
  reactions.   This proximity of scales
  indicates that the corresponding potentials are
  actually too  soft for the purpose of applications to the pion
  production reaction.  Even for the hardest available choice of the
regulator in the new chiral $NN$ potentials, one may expect that a significant portion of the dynamical
intermediate-range physics is effectively transferred from the NNLO amplitude
to the NNLO contact term, which is thus   enhanced compared to
the evaluation in the previous sections.
In order to explicitly show the effects of such low-momentum cutoffs,
we  calculate the pion-production amplitude using the convolution of the operators from
Sec.~\ref{sec:operators}
with chiral $NN$ wave functions.

\begin{figure}[t]
  \centering
  \begin{minipage}[b]{0.45\textwidth}
    \includegraphics[width=\textwidth]{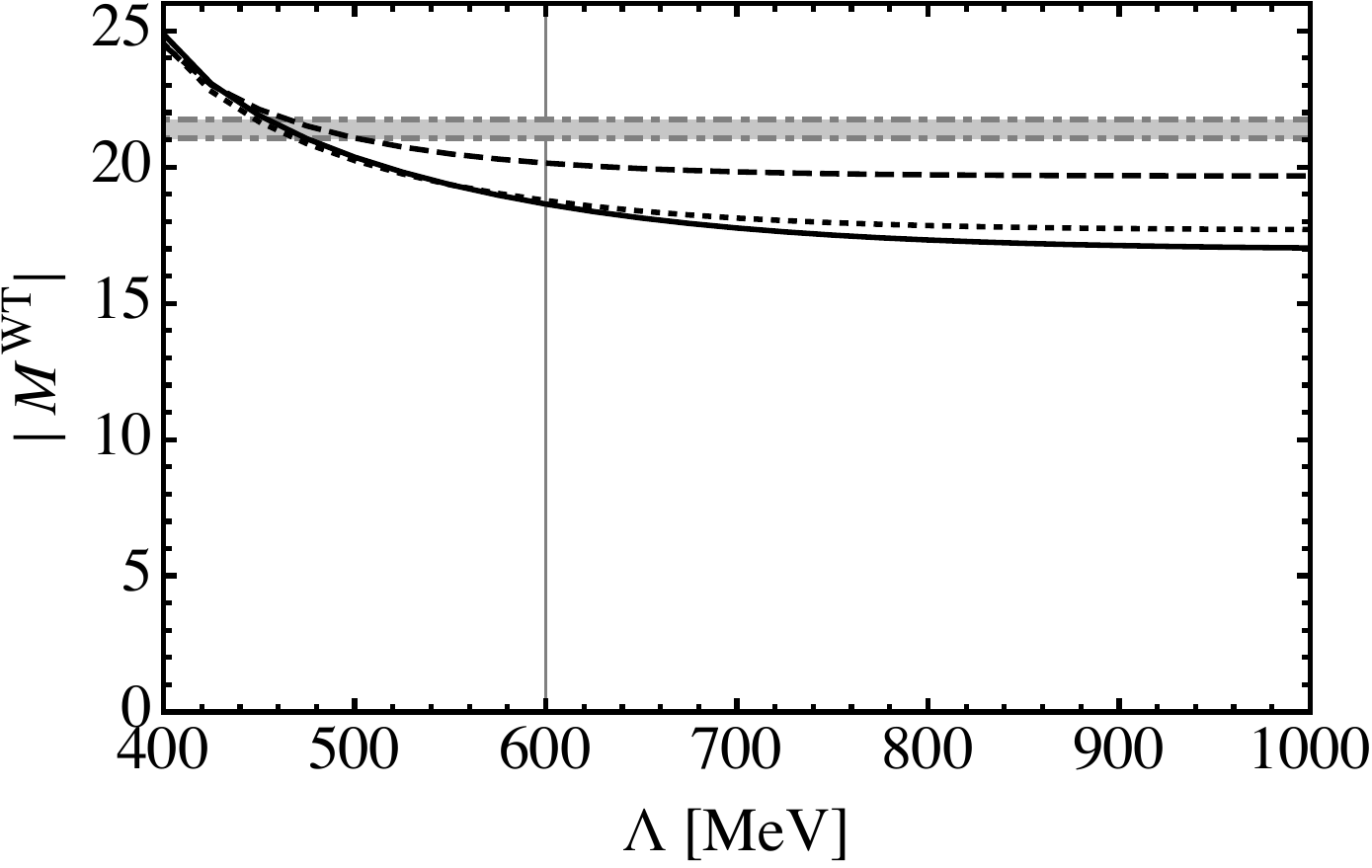}
  \end{minipage}
  % \hfill
  \qquad
  \begin{minipage}[b]{0.45\textwidth}
    \includegraphics[width=\textwidth]{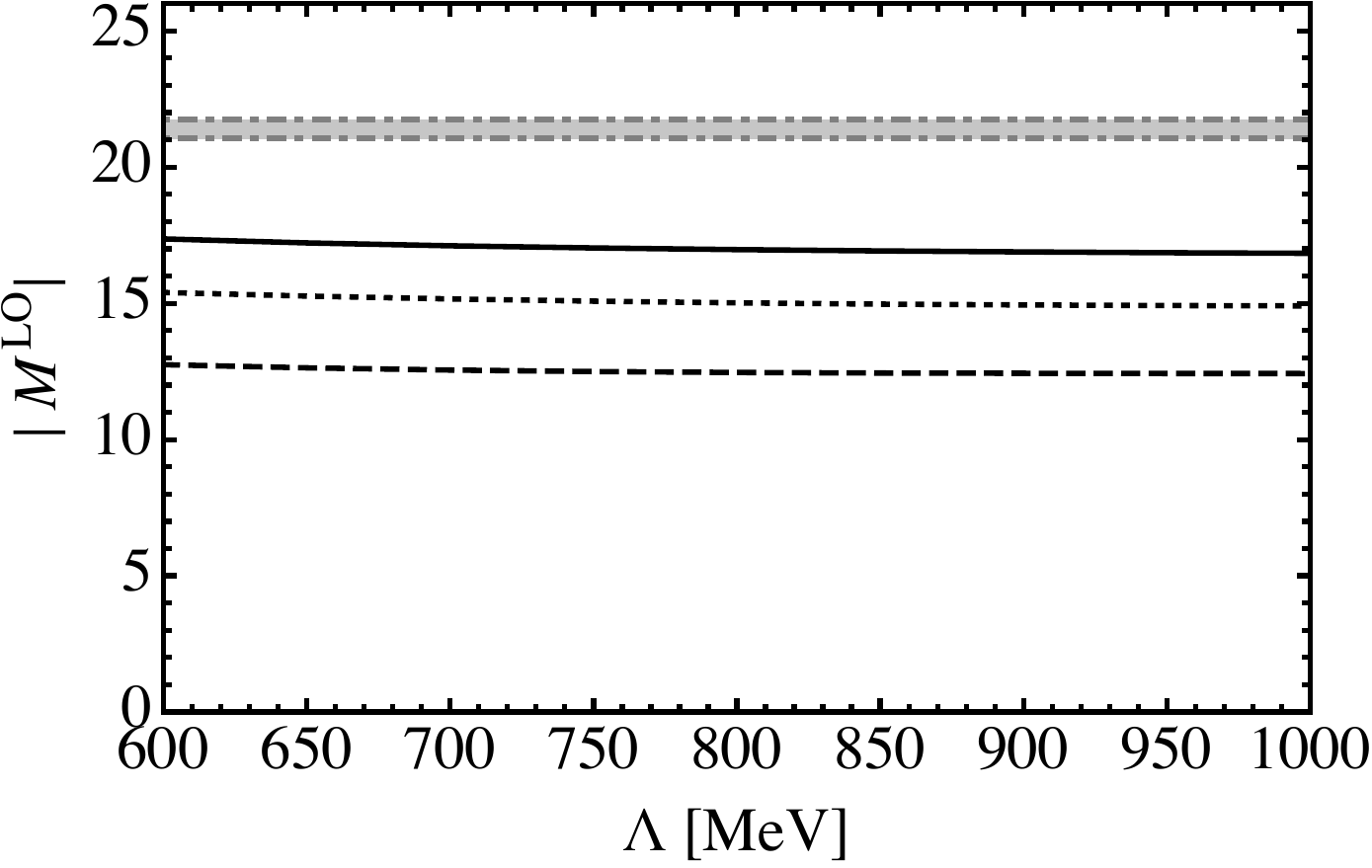}
  \end{minipage}
  \caption{\label{fig:amplWTandLOchiral}
  The $pp \to d\pi^+$ amplitude $|M_{3P1}|$ based solely on
  the LO Weinberg-Tomozawa operator (left panel)
  and all LO operators (right panel) as a function of the sharp
  momentum integral cutoff $\Lambda$. The amplitude is calculated
  using the
  \emph{chiral} $NN$ wave functions at N$^4$LO for
  different choices of the regulator, namely $R=0.8$~fm (solid orange
  line),  $R=1.0$~fm (dotted orange
  line) and $R=1.2$~fm (dashed orange
  line).
See    Fig.~\ref{fig:amplWTandLO} for notation  of
    the horizontal  grey band.
  }
\end{figure}

We again first consider the long-range LO WT contribution to the pion-production operator.
The result of its convolution with chiral wave functions is shown in
Fig.~\ref{fig:amplWTandLOchiral} (left).
As expected, the amplitudes are very similar to the ones obtained with
phenomenological potentials (cf.~Fig.~\ref{fig:amplWTandLO} left).
We next consider the complete LO operator including the direct term.
The inclusion of the LO direct pion production operator yields the amplitude
shown in the right panel of Fig.~\ref{fig:amplWTandLOchiral}.
We see that the contribution of the direct operator is no longer small
(compared to the calculation with phenomenological
$NN$ wave functions Fig.~\ref{fig:amplNNLO}).
Furthermore, as expected, the inclusion of the direct term generates a
dependence
on the short-range
details of the chiral $NN$ potentials.
One can clearly see the pattern: chiral potentials with higher momentum-space
cutoffs produce results closer to the ones based on (harder)
phenomenological potentials and also to experimental data.
This pattern is to be expected and provides an illustration
of how a  part of the intermediate-range contribution to the
amplitude is reshuffled into the contact interaction upon explicitly
integrating out the momentum components of the nucleons above the scale
$\Lambda_{NN}$.
The result for the chiral potential with a cutoff $R=0.8$ fm is  rather close to
the result using AV18,
cf.~the left panels of Figs.~\ref{fig:amplWTandLO} and~\ref{fig:amplWTandLOchiral}.
Finally, we remark that our results using the complete set of operators
up-to-and-including NNLO  are similar to the ones at LO.\@
In other words,
the inclusion of all LO, NLO and NNLO terms does not change the chiral $NN$ potential
cutoff pattern displayed in Fig.~\ref{fig:amplWTandLOchiral}
right.

\begin{figure}[t]
  \centering
  \includegraphics[width=.45\textwidth]{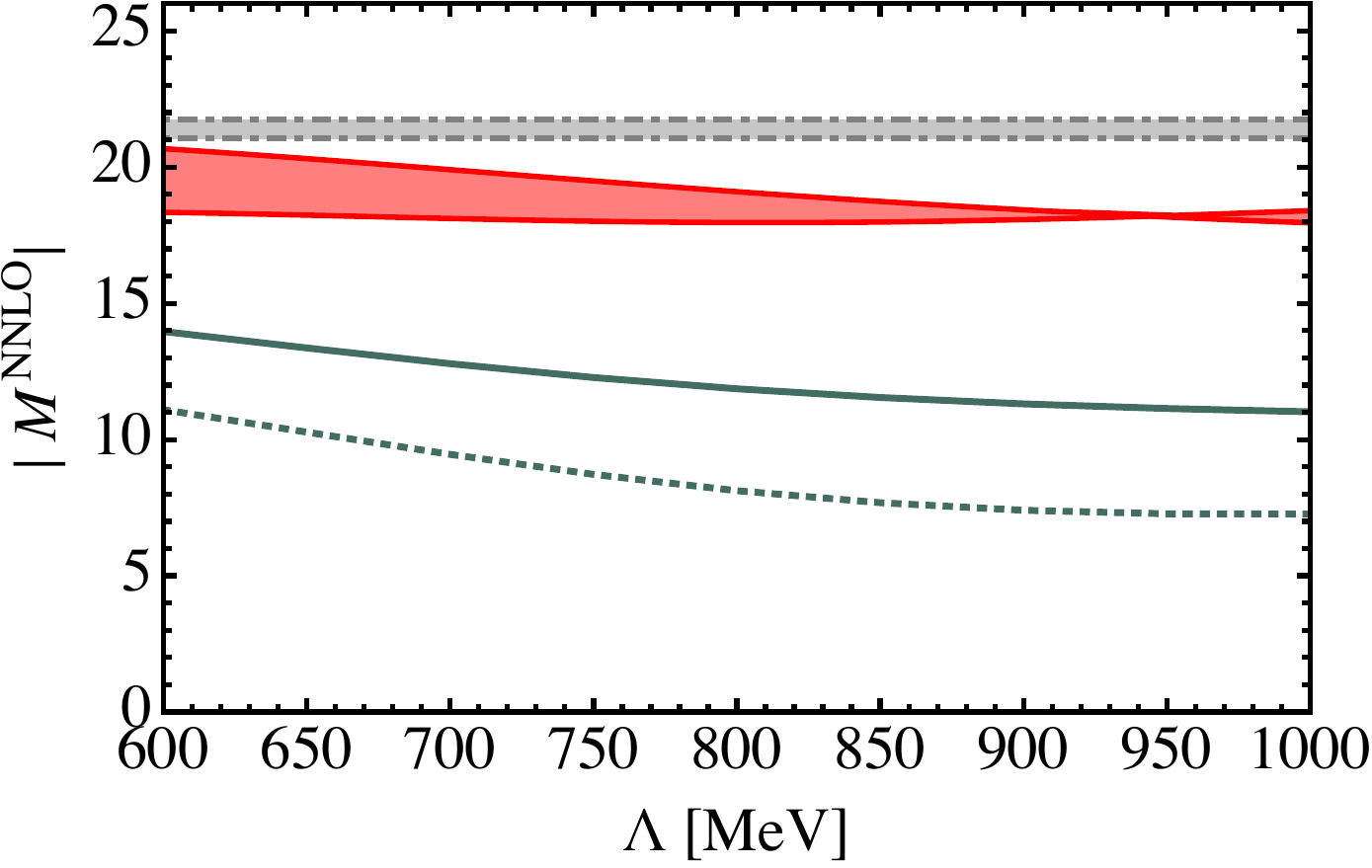} 
  \caption{\label{fig:amplchiral3and4}
    The $pp \to d\pi^+$ amplitude $|M_{3P1}|$ from
    the complete set of LO, NLO and  NNLO operators
    \underline{without} any contact term  in the $NN\to NN\pi$ transition operator
    as a function of the sharp
  momentum integral cutoff $\Lambda$ for the $NN$ potentials at
  N$^3$LO~\cite{Epelbaum:2014efa}
  (dotted line) and N$^4$LO~\cite{Epelbaum:2014sza}
  (solid line) corresponding to the choice of the regulator of
  $R=0.9$~fm.    The (red)  band  is the same as  in Fig.~\ref{fig:amplNNLO-ciValues},
 for notation  of  the horizontal  grey band see  Fig.~\ref{fig:amplWTandLO}.
  }
\end{figure}

Finally, in Fig.~\ref{fig:amplchiral3and4} we plot the $|M_{3P1}|$ amplitude, where the
$NN$ initial and final state wave functions are generated based on the
N$^3$LO~\cite{Epelbaum:2014efa}  and N$^4$LO~\cite{Epelbaum:2014sza} $NN$ potentials, with the
regulator $R=0.9$~fm, which was found to yield the smallest theoretical uncertainties for
$NN$ observables. Interestingly, we observe a significant sensitivity of
the
calculated amplitude to the intermediate-range components of the $NN$
potential.\footnote{In addition to the isospin-breaking contact
  interaction in the $^1$S$_0$ channel, the only new ingredient in the
  $NN$ potential at N$^4$LO of Ref.~\cite{Epelbaum:2014sza} is given
  by the corresponding (parameter-free) two-pion exchange
  contributions.}  We emphasize, that the results obtained using   the N$^4$LO chiral  $NN$ wave functions  lie  closer to
 the experimental data and therefore yield more natural  values for the NNLO contact term contribution.
The difference between the two results is comparable in size
to the cutoff variation discussed above.

%%%%%%%%%%%%%%%%%%%%%%%%%%%%%%%%%%%%%%%%%%%%%%%%%%%
\section{Summary}
\label{sec:concl}
%%%%%%%%%%%%%%%%%%%%%%%%%%%%%%%%%%%%%%%%%%%%%%%%%%%

Pion production in $pp \to d\pi^+$ reaction is studied at threshold within chiral EFT.\@
Using a  complete set of  pion production operators  derived in Refs.~\cite{Filin:2012za,Filin:2013uma}
up-to-and-including next-to-next-to-leading order (NNLO)
for s-wave pions and a  set  of modern phenomenological and
  chiral $NN$ potentials   we calculate the threshold observable,  namely the absolute value of the $pp \to d\pi^+$ reaction amplitude,
  and compare it to the experimental data.  We emphasize that up to next-to-leading order  our results
  are parameter free  while at NNLO there is one  unknown $NN\to NN\pi$ contact term.
  Apart from the description of data,   the goal of this study was to  demonstrate that the momentum counting scheme (MCS)
  used to classify the operators is  adequate  and that the theoretical  uncertainty can be estimated
  reliably based on our expansion parameter.  In particular,   our  results at NNLO  serve to comply  with this goal.
 Another   goal of this work was  to incorporate the $\Delta(1232)$  resonance in the  analysis in order to  investigate
 its  role as an explicit  degree of freedom for  pion production reactions.
 Furthermore,  we  studied the sensitivity of the  results  to  various $NN$ wave functions.

 As  is known from the previous studies,   the results in the $pp \to d\pi^+$ channel  are governed  by the longest range
 Weinberg-Tomozawa  operator  at   leading order (LO) (see the top row  in Fig.\ref{fig:allN2LO})
 which alone yields the amplitude comparable to the experimental data.  This result appears to be nearly
 independent of the  $NN$ model used.
 The contributions at   next-to-leading order (NLO) undergo significant cancellations:   while most of  the
 loop contributions including pions, nucleons and $\Delta$ vanish exactly at this order,
 destructive interference of the  diagrams which possess a  (reducible) $N\Delta$ intermediate state
 (see diagrams dir$\Delta a$ and $\Delta $Box\,$a$ in Fig.\ref{fig:allN2LO}),
 is not exact yielding a finite but  small  contribution comparable in  size with the  NNLO corrections.
We observe that   
the available chiral potentials are generated with a cutoff,
which tends to remove a part of the intermediate range physics relevant for  the reaction $NN \to NN\pi$.
On the other hand, we demonstrated that when the $NN$ wave functions 
 are calculated based on phenomenological potentials, which by construction take  
momenta significantly larger than the momentum scale inherent for pion production in $NN \to NN\pi$,
the size of the contributions at LO, NLO and NNLO  turns out to be in agreement with the expectations of the momentum counting scheme.
Further,  we find that the  variation in the NNLO  results due to the use of different  $\pi N$ low-energy
constants (LECs) $c_i$  is consistent with  the uncertainty estimate expected at  NNLO.
In addition,  some higher-order corrections  from  the nucleon recoil terms  in the $\pi NN$ propagators  
of the rescattering operators were evaluated 
 explicitly to confirm that they are fully in line with the  MCS estimate. \@

Apart from the state-of-the-art calculation of
$pp \to d\pi^+$ reaction, the  results of this work provide  an important step towards
a quantitative understanding of  the much more challenging   $pp \to pp\pi^0$ channel.  The
 consistency of the power counting  verified  with the explicit calculations presented in this work
 (at least for phenomenological $NN$ wave functions)  is a necessary pre-requisite for studying
the chirally suppressed  neutral pion production.

Finally,   the role of the  $\Delta$  resonance was studied   using three  different strategies:
 (i)  the most complete formulation  with   $\Delta$-excitations being included in the Hilbert
 space which results in a coupled-channel framework (CC$\chi$EFT$-\Delta$) 
 (ii)  a perturbative treatment  of  $\Delta$,  where   all effects
associated with the $\Delta$-nucleon mass difference   are included in the pion production operator  while
the Hilbert space consists of only nucleonic states ($\chi$EFT$-\Delta$);
(iii)  $\Delta$-less formulation of chiral EFT ($\chi$EFT$-\Delta \palka$)
where all effects due to the $\Delta$ isobar are integrated out and included in the  LECs  $c_i$.

We find that   the difference between  the  approaches  
CC$\chi$EFT$-\Delta$  and $\chi$EFT$-\Delta$  does not exceed the magnitude of  NNLO  effects, if the  cutoff 
is chosen in the range,  which allows for a sufficient separation of the soft and hard scales ($\Lambda\ge 700$ MeV).
 Furthermore, the difference between the  $\Delta$-full and  $\Delta$-less  approaches,  
  which is  expected to  be of the size of the NLO corrections,  is also comparable with  the NNLO estimate.
   The smaller than expected difference
 can be attributed to    destructive  interference   between the individually sizeable  diagrams  at NLO
involving  a (reducible) $N\Delta$ intermediate state.

\section{Acknowledgments}

We  thank Arnoldas Deltuva for providing us with the amplitudes generated
from the CD Bonn  $NN$ model with explicit $\Delta$'s. This work was supported by the DFG (SFB/TR 16, ``Subnuclear Structure of Matter''), 
the European Research Council (ERC-2010-StG 259218 NuclearEFT)
and the National Science Foundation Grant No.~PHY-1068305.
This work was partially supported through GAUSTEQ (Germany and U.S. Nuclear Theory Exchange Program for QCD Studies of Hadrons and Nuclei) under contract number DE-SC0006758.  One of the authors (V.B.)  thanks  the DFG (grant GZ: BA 5443/1-1 AOBJ:  616443)  for the partial support.

% \appendix

\appendix
\section{Pion production operator up to  NNLO}
\label{app:operators}

In this  appendix we   list,    for the sake of completeness,  the  operators
 for  s-wave  pion  production from  the diagrams
shown  in Fig.~\ref{fig:allN2LO}   up-to-and-including  NNLO  in the
$\chi$EFT$-\Delta$ framework.
Here  we  present the  renormalized result for the threshold production operators.
Details of  the renormalization procedure can be found in
Refs.~\cite{Filin:2012za,Filin:2013uma},  where these operators
were derived.

In the MCS, the rescattering operator  at  LO  involves  the Weinberg-Tomozawa $\pi N$
 vertex  (including its recoil correction)    which  yields 
 \begin{eqnarray}
	i M^\text{LO}_{\rm rescat}  
	=\frac{g_A\, (2m_N)^2}{4 f_\pi^3} 
	\frac{ m_{\pi}}{2\omega(k)} \left(\frac1{P_1}+\frac1{P_2}\right)
		(\vec\sigma_2 \cdot \vec k) \taux^a \, + (1 \leftrightarrow 2),
\label{eq:treeLO}
\end{eqnarray}
where $f_\pi =92.4$ MeV is the pion decay constant,  $g_A=1.32$ is the axial constant,  
 and  the pion propagator is written in terms of  time-order-perturbation-theory (TOPT) (for details see Eqs.\eqref{topt} below). 
 Here,    $\taux^a$  is the
antisymmetric   isospin operator,
$\taux^a  = i (\boldtau_1 \times \boldtau_2)^a$,
with  the superscript $a$ ($a$=1,2,3)  referring  to the isospin quantum number of  the outgoing pion field,
 $\vec \sigma$ is a three-vector of Pauli matrices   and $k_i=p_i-p_i'$,  where
$p_i$ ($p_i'$)  stands for the momentum of the initial (final) nucleon
$i$ ($i=1,2$).  
  The nucleon bispinors  are normalized as   $\bar u u =2m_N$ 
 which accounts for  the appearance of a  factor  $(2m_N)^2$  in the amplitude \eqref{eq:treeLO} 
and  similar factors in the other amplitudes below.  We note further that    in the center of mass system  the  kinematics relevant for threshold pion production reads
 \begin{eqnarray}
   \vec p_1 = -\vec p_2  \eqqcolon \vec p,  \quad
    {\vec p_1}^{\,\prime} = -{\vec p_2}^{\,\prime}   \eqqcolon {\vec p}^{\,\prime}, \quad  \vec k_1=-\vec k_2   \eqqcolon \vec k\,.
\end{eqnarray}
The interchange  $(1 \leftrightarrow 2)$  in Eq.~\eqref{eq:treeLO} (and the expressions below) indicates 
that the permutations of the initial and final nucleons  need  to be included (one needs also to take into account that $\vec k\to -\vec k$ under this interchange). 

 The  tree-level rescattering operator  at  NNLO  is decomposed into two parts:
 the first part, $M_{\rm rescat1}^\text{NNLO}$,
 contains   the  corrections
suppressed as  $1/m_N$  due to  the vertices  from ${\cal L}^{(2)}_{\pi\!N}  $
while the second term,  $M_{\rm rescat2}^\text{NNLO}$,
accounts for the corrections  $\propto 1/m_N^2$  from ${\cal L}^{(3)}_{\pi\!N}$.

The  explicit TOPT expressions  read  
%%%%%%%%%%%%%%%%
\begin{eqnarray}
	i M_{\rm rescat1}^\text{NNLO} &=& \frac{g_A\, (2m_N)^2}{2f_\pi^3}    \tau_2^a  (\vec\sigma_2 \cdot \vec k)
		\left[
			 \frac{4c_1 m_\pi^2 }{2\omega(k)} \left(\frac1{P_1}+\frac1{P_2}\right) -
                         \left( 2c_2+2c_3-\frac{g_A^2}{4 m_N} \right) \frac{ m_{\pi} }2 \left(\frac1{P_1}-\frac1{P_2}\right) 
		\right]
\nonumber
\\
            &-& \frac{g_A\, (2m_N)^2}{2f_\pi^3}
	{m_{\pi}\,  \taux^a } \,
           \frac{\vec\sigma_2 \cdot (\vec p +{\vec p}^{\,\prime})}{4 m_N}   \frac{ 1 }2 \left(\frac1{P_1}-\frac1{P_2}\right) 
            +   (1 \leftrightarrow 2),
\label{treeNNLO1}
\\
	i M_{\rm rescat2}^\text{NNLO} &=&
	-\frac{g_A(2m_N)^2}{2f_\pi^3}
	 \frac{m_{\pi}}{2\omega(k)} \left(\frac1{P_1}+\frac1{P_2}\right) 
	 	\Bigg\{
	\tau_2^a \,  (\vec\sigma_2 \cdot \vec k) \,
	  \frac{\vec{p}^{\,2}-\vec{p}^{\,\prime 2}}{m_N^2} \left( m_N c_2 -\frac{g_A^2}{16} \right)
\nonumber
\\
	&-& \taux^a \, (\vec\sigma_2 \cdot \vec k) 
		\left[
			\frac{\vec{p}^{\,2}+\vec{p}^{\,\prime 2}}{16 m_N^2}
			-\left( \frac{1+g_A^2+8m_N c_4}{8 m_N^2} \right)
                 \left( i \vec\sigma_1 \cdot  (\vec p \times \vec p^{\, \prime})
                   - \frac{ m_{\pi}^2}{2}\right)
		\right]
\nonumber
\\	\nonumber &-& \frac{\taux^a }{8 m_N^2}
		\left[
			(\vec\sigma_2 \cdot {\vec p}^{\: \prime}) \vec p^{\, 2} -  (\vec\sigma_2 \cdot \vec p )\vec p^{\,\prime 2}
		\right]\Bigg\}\\
		&+&\frac{g_A(2m_N)^2}{4f_\pi^3}
	 m_{\pi}  	\taux^a \, (\vec\sigma_2 \cdot \vec k) 
			 \frac{1+g_A^2+8m_N c_4}{8 m_N^2} 		 +  (1 \leftrightarrow 2).
\label{treeNNLO2}
\end{eqnarray}
 
 To arrive at  the expressions \eqref{eq:treeLO},  \eqref{treeNNLO1} and  \eqref{treeNNLO2},  we used that  the   pion propagator in TOPT  reads
\begin{eqnarray}\label{topt} 
\frac1{k_2^2-m_\pi^2+i0} &=& \frac1{2\omega(k)} \left(\frac1{P_1}+\frac1{P_2}\right),\quad \quad
\frac{v \cdot k_2}{k_2^2-m_\pi^2+i0} =  \frac12 \left(\frac1{P_1}-\frac1{P_2}\right),\\
P_1&=&\sqrt{s} - 2m_N - \frac{\vec p^{\, 2}}{2m_N} - \frac{\vec p^{\,\prime 2}}{2m_N}- \omega(k),\\   
P_2&=&\sqrt{s}  - 2m_N-m_{\pi} - \frac{\vec p^{\,  2}}{2m_N} - \frac{\vec p^{\,\prime 2}}{2m_N}  - \omega(k) ,
\end{eqnarray}
where  $\sqrt{s}=   m_d+m_{\pi}$,  $\omega(k)=\sqrt{m_{\pi}^2+\vec k^2}$  and $v \cdot k_2$ stands for the zeroth component of the four-vector $k_2$.
Note that the leading effect  from the propagators stems  from the pion three-momentum squared,  whereas the nucleon recoils  are suppressed by two  orders in the MCS. 
In order to retain all terms at  NNLO,  one therefore needs to keep the recoil terms  in  the  leading Weinberg-Tomozawa operator \eqref{eq:treeLO}.  
Meanwhile, in the operators \eqref{treeNNLO1}  and  \eqref{treeNNLO2}, which  start  to contribute at NNLO, it suffices to preserve only the leading term in the propagators.  However, 
we  retain the recoil  terms  in the evaluations  of these operators to maintain the   correct analytic structure of the three-body propagators and to  have an estimate  of higher order terms.  
We find that  the  combined effect from all  the recoil terms in Eqs.\eqref{treeNNLO1} and  \eqref{treeNNLO2}  is about  2\%  of the leading order amplitude, in full agreement with 
the estimate of N$^4$LO  corrections. 

Furthermore,  we emphasize that   the very last term in Eq.~\eqref{treeNNLO2}  has the  form which coincides  exactly  with the structure of the  NNLO $NN\to NN\pi$ contact operator (see 
  diagram in the last row in Fig.~\ref{fig:allN2LO}). Indeed,  the contact term contribution for s-wave pion production in   $pp\to d \pi^+$ reaction channel  can be written as
\begin{eqnarray}\label{MCT}
i M_{\rm CT}^\text{NNLO} =  \frac{ (2m_N)^2}{  f_\pi^3} 
	 { m_{\pi}} \, C \,   
		(\vec\sigma_2 \cdot \vec k) \taux^a \, + (1 \leftrightarrow 2),
\end{eqnarray}
where the renormalized part of  the LEC $C$  is of  the order of  $\Lambda_{\chi}^{-2}$ while its divergent part cancels the divergent terms  from the NNLO loop contributions,
 as discussed in  Refs.~\cite{Filin:2012za,Filin:2013uma} (see also a review article \cite{Baru:2013zpa}).  
Since the details of the short-range mechanisms can not be revealed in an EFT study,  it is  convenient to absorb the 
last term in  Eq.~\eqref{treeNNLO2} into the redefinition of the LEC $C$.   The results presented in  this paper are therefore obtained using this formulation.  
We note, however,  that  keeping  the last term in Eq.~\eqref{treeNNLO2} explicitly  is equally justified and  does not affect the conclusions about 
the applicability of the MCS power counting drawn in this paper.  For the results obtained in the formulation where the   last term in Eq.~\eqref{treeNNLO2} 
 is retained, the interested reader is referred to Ref.~\cite{Filin_CDtalk}.
 
In  addition to the rescattering operators,  there are the so-called  direct  diagrams
which   respond for    the  direct pion  emission from  a single nucleon,   see
the first diagram in the first (fifth) row of Fig.~\ref{fig:allN2LO},   which contributes  at  LO   (NNLO).

The  contribution  of   the  ``direct''  diagrams to the pion production operator  is
a one-nucleon operator and can be  written as
%%%%%%%%%%%%%%%
\begin{eqnarray}
	i M_{\rm dir}& =& \frac{g_A\, (2m_N)}{f_\pi} \tau_1^a  \, m_{\pi}
            (2\pi)^3\delta(\vec p- {{\vec p}}^{\, \prime}) 
%\nonumber
\\
&&\times
	\left[ \frac{1}{4 m_N}  \vec\sigma_1   \cdot (\vec p+\vec p^{\, \prime})
		- \frac{1}{16 m_N^3} \left(   {\vec p}^{\,2} ( \vec\sigma_1  \cdot \vec p)
		+   \vec p^{\,\prime 2} ( \vec\sigma_1  \cdot \vec p^{\, \prime})\right)
	\right] +   (1 \leftrightarrow 2)\,.
\nonumber
\label{Mdir}
\end{eqnarray}
%%%%%%%%%%%%%
As explained in Ref.~\cite{Filin:2013uma}, this  amplitude   contributes to observables  only  when  convoluted  with the initial and final
$NN$ wave  functions.

Further, the pion production operator contains diagrams with the  intermediate $N\Delta $ state shown in the second row  in Fig.~\ref{fig:allN2LO}.  
The diagrams   ``${\rm Dir \Delta a}$''  and  the   box   diagram  ``$\Delta$Box a''  give rise to the contributions relevant for the $pp\to d\pi^+$ channel while all four diagrams 
contribute to $pp\to pp\pi^0$. 
These diagrams  should  be added to the operator derived  in Ref.~\cite{Filin:2013uma}   if the  $\Delta$  contributes   
as a  part  of the $NN\to NN\pi $  operator, i.e.~in the $\chi$EFT$-\Delta$ framework.
 On the other hand,   when the  $N\Delta $  state  is  included  in  the coupled-channel formalism  the contributions  of these diagrams are  generated
 automatically  when the  $N\Delta\to NN\pi$ ($NN\to N\Delta\pi$) operators shown in
 Fig.~\ref{fig:treeDelta}  are convolved  with the  initial state $NN\to
 N\Delta$ (final state $N\Delta\to NN$) amplitude.
The diagrams in the second row  in Fig.~\ref{fig:allN2LO}, therefore, should be omitted in the CC$\chi$EFT$-\Delta$ approach.

The expression for  the  ``direct''  diagram Dir$\Delta$a      in Fig.~\ref{fig:allN2LO}  reads  
\begin{eqnarray}\nonumber\label{DirDeltaa}
	i M_{\rm Dir \Delta a}^{\rm NLO} &=& -\frac{\ga\, \gpind^2 \, (2m_N)^2}{2\fpi^3} \frac{\mpi}{\mn}   (\vec \sigma_2\cdot \vec k)
		\left( \frac{2}{3} (\vec p^{\,\prime} \cdot \vec p - \vec p^{\,\prime 2}) - \frac{i}{3} \vec \sigma_1 \cdot (\vec p^{\,\prime} \times \vec p) \right)
		\left( \frac{1}{3} \tau_\times^a + \frac{2}{3} \tau_2^a \right) \\
		&& \times \frac{1}{\sqrt{s}  - 2m_N - \delta - \frac{\vec p^{\,\prime 2}}{\mn} }\;
		\frac{1}{\vec k^2 +m_{\pi}^2}
		 +  (1 \leftrightarrow 2),
\end{eqnarray}
where     $\delta = m_\Delta - m_N$, 
the first  propagator stands for the TOPT propagator corresponding to the $N\Delta$  intermediate state while
 the second one corresponds  to the static   OPE propagator in the $NN\to N\Delta$ transition. 
 Although   the term $\sim m_\pi^2$  in the OPE    propagator gives rise to higher-order effects at   
next-to-next-to-next-to-leading order (NNNLO),  we   keep it  to allow for  a close  comparison  with 
 our results  based on a  coupled-channel  approach (see Sec.~\ref{sec:delta} for a  detailed discussion of the results).  
 In particular, in order to obtain the results in the  coupled-channel framework 
 (CC$\chi$EFT$-\Delta$) we utilized the    $NN\to N\Delta$ transition amplitude   generated 
 in  Ref.~\cite{Deltuva:2003wm} using iterations of the static OPE potential.   
 We checked that  neglecting  the  term $\sim m_\pi^2$ in the OPE propagator results in  an about 3\%  correction  
to the  reaction amplitude which is fully in line with NNNLO estimate.  
Further,      for the purpose of study $NN \to NN \pi$ reaction near threshold   to the order we are working,
  the  width of the $\Delta$-resonance in the propagators can be safely neglected:
  while the $\Delta$ width  vanishes exactly at    pion production threshold, it  constitutes a higher order effect for the energies near threshold. 
 
The  expression for the    box   diagrams  ``$\Delta$Box a''  and ``$\Delta$Box b''  (the second row  in Fig.~\ref{fig:allN2LO}) is
\begin{eqnarray}\nonumber\label{DeltaBoxa}
	i M^{\rm NLO}_{\Delta \text{box}} &=&  \frac{ \ga\gpind^2\, (2m_N)^2}{36 \fpi^5} \; m_{\pi} \;
		\bigg[
			 3\tau_{+}^a \; 
			 i \vec k \cdot (\vec \sigma_1 \times \vec \sigma_2)
			\left( I_{\rm sum} - \frac14\frac{\vec k^2}{\delta} (  \jpipid +   J_{\pi \pi N}) \right) \\
			&&-  (\vec \sigma_1+\vec \sigma_2)\cdot \vec k \;  \tau_\times^a
				\left( I_{\rm sum} - \frac12\frac{\vec k^2}{\delta}(  \jpipid +   J_{\pi \pi N}) \right)
		\bigg] +  (1 \leftrightarrow 2),
\end{eqnarray}
where $g_{\pi N\Delta} =1.34$   is the leading $\pi N \Delta$ coupling constant, $\taup^a$ is the symmetric isospin operator, $\taup^a = (\boldtau_1+\boldtau_2)^a$, and   the integral  combination  $I_{\rm sum}$ is
\begin{eqnarray}\label{Isum}
	I_{\rm sum} = \ipipi + \frac12 \frac{\jpid}{\delta} + \delta \jpipid + \frac{2}{(4 \pi)^2}.
\end{eqnarray}
The individual  integrals   
are defined below,  see Eqs.~(\ref{JpiD}--\ref{JpipiND}).
Note that the integrals $I_{\rm sum}, \jpipid$   and $J_{\pi \pi N}$ are finite ($I_{\rm sum}$ and  $\frac{1}{\delta} (  \jpipid +   J_{\pi \pi N})$   vanish  in the limit $\delta \to \infty$)  
while the divergent part of the loop is absorbed in the $NN\to NN\pi$ contact term contribution at NNLO, see Eq.~\eqref{MCT}.

The  contribution  of  pion-nucleon  loops  to  the  production  operator   for
s-wave  pions  was  derived in   Ref.~\cite{Filin:2012za}.  After renormalization the finite part  of  these loops reads   
\begin{eqnarray}\nonumber
i M^{\text{\NNLO{}}}_{\pi N\rm-loops} &=& -\frac{\gA\, (2m_N)^2 \ m_{\pi}}{4\fpi^5}
\taux^a  (\vec\sigma_1 + \vec\sigma_2)\cdot \vec k
\left[ \frac16 I^R_{\pi\pi}(k_1^2) \left(1-\frac{19}{4}g_A^2 \right)-
\frac{1}{18(4 \pi)^2 } \left(1-10g_A^2 \right) \right] \; \\
    &-&i\,\frac{\gA^3\, (2m_N)^2 \ m_{\pi}}{4\fpi^5}
    \taup^a \, 
     \vec k \cdot (\vec \sigma_1 \times \vec \sigma_2)
     \, I^R_{\pi\pi}(k_1^2) +  (1 \leftrightarrow 2),
\label{eq:Mga3final}
\end{eqnarray}
where 
the integral   $I_{\pi\pi}^R(k_1^2)$  is  defined  below, see
Eq.~(\ref{ipipifsi})\footnote{
%%%%%%%%%%
In Ref.~\cite{Filin:2012za},
the  integral $\ipipi(k_1^2)$  was  called  $J(k_1^2)$.
%%%%%%%%%
}.

Finally,   the  renormalized, finite  $\Delta$ loop diagrams contribution to
s-wave pion  production at  NNLO  reads:
%%%%%%%%
\begin{eqnarray}
\label{delsymmampstart}
\nonumber
		 \hspace*{-0.5cm} i M_{\Delta\text{-loops}}^{\text{\NNLO{}}}
       &=&
      	  - \hspace*{-0.cm}      \frac{\ga \gpind^2\, (2m_N)^2 }{4\fpi^5}\; m_{\pi} \, \tau_{\times}
        \;  (\vec  \sigma_1+  \vec \sigma_2) \cdot \vec  k
   \\
      \nonumber\label{delsymmampend}
      &\times&\hspace*{-0.cm}
        \Bigg[
 	    	\frac{5}{9}I_{\rm sum} 
	    	- \frac{1}{18} {\vec k}^2 \jpipind
			     - \frac{8}{9} \frac{\delta^2}{\vec k^2} 
	I_{\rm sum}  	- \frac{2}{27} \left(  \intI +
          \frac12 \intJ + \frac13 \intC{2} \right)
	        \Bigg]
	          \\
%           \nonumber
     &+&  i \frac{\ga \gpind^2\, (2m_N)^2}{8\fpi^5} m_{\pi} \,
	    \tau_{+}^a	\;    
	      \vec k \cdot (\vec \sigma_1 \times \vec \sigma_2)
	    \Bigg[
	    	    \frac29 I_{\rm sum} 
	    	  - \frac{1}{18} {\vec k}^2 \jpipind
	    \Bigg]
	         +  (1 \leftrightarrow 2).
\end{eqnarray}
The   dimensionless loop integrals entering Eqs.~\eqref{DeltaBoxa},  (\ref{eq:Mga3final}) and (\ref{delsymmampstart}) are  defined as follows
\begin{eqnarray}
	\frac{1}{\delta}\jpid (\delta)
		   &=& \frac{\mu^\epsilon}{i \delta} \int \frac{d^{4- \epsilon}l}{(2\pi)^{4-\epsilon}}
	         \frac{1}{(l^2-\mpi^2+i0)( - v \cdot l - \delta +i0)}	\,,
		\label{JpiD}
\\
	\ipipi (k_1^2)
		&=& \frac{\mu^\epsilon}{i} \int \frac{d^{4- \epsilon}l}{(2\pi)^{4-\epsilon}}
		\frac{1}{(l^2-\mpi^2+i0)((l+k_1)^2 - \mpi^2 +i0)} \,,\\
		\label{Ipipi}
	\intT(k_1^2,\delta) &=& \delta \frac{\mu^\epsilon}{i} \int \frac{d^{4- \epsilon}l}{(2\pi)^{4-\epsilon}}
         \frac{1}{(l^2-\mpi^2+i0)((l+k_1)^2 - \mpi^2 +i0)( - v \cdot l - \delta +i0)},\phantom{MM}
         \label{JpipiD}
\\
	{\vec k}^2 J_{\pi \pi N \Delta}(k_1^2,\delta)&=&\frac{{\vec k}^2 }{\delta}(  \jpipid -   J_{\pi \pi N}),
         \label{JpipiND}
\end{eqnarray}
where $J_{\pi \pi N}(k_1^2)=J_{\pi \pi \Delta}(k_1^2,\delta=0)$.
The integrals (\ref{JpipiD}) and (\ref{JpipiND}) as well as the linear combination $I_{\rm sum}$ from Eq.~\eqref{Isum}  are finite and were evaluated numerically, while
the integrals $\jpid$ and $\ipipi$  contain finite and  divergent parts.
The renormalized finite parts  of $\jpid$ and $\ipipi$ are given by~\cite{Filin:2013uma}:
%%%%%%%%%%%%%%%%%%%%%%%%%%%%%%%%%%%%%%%%%%%%%%%%%%%%%%%%%%%%%%%
\begin{eqnarray}
   \hspace*{-1cm}   I^R_{\pi\pi}
        &=&   - \frac{1}{(4\pi)^2} \log \left( \frac{\mpi^2}{\mu^2} \right) +		\frac{1}{(4\pi)^2}
		\left(
		1-  2 \frac{\sqrt{4-x-i0}}{\sqrt{x}} \arctan \left( \frac{\sqrt{x}}{\sqrt{4-x-i0}} \right)
		\right),
		\label{ipipifsi}
\\
 \hspace*{-0.92cm}   \frac{1}{\delta}J^R_{\pi\Delta}
           \!&\!=\!&     \!         \frac{2}{(4\pi)^2}
           \log \left( \frac{\mpi^2}{\mu^2} \right)   \!+\!           \frac{4}{(4\pi)^2}
               \left\{ -\frac{1}{2} +
                    \frac{\sqrt{1-y\!-\!i0}}{\sqrt{y}}
                    \left[
                          \!-\frac{\pi}{2} \!+ \arctan \left( \frac{\sqrt{y}}{\sqrt{1-y\!-\!i0}} \right)
                    \right]
               \right \}\!\!,
\end{eqnarray}
where $\mu$  is the  dimension-regularization scale which is chosen to be
$\mu \simeq 4\pi \fpi (\simeq \Lambda_\chi \simeq m_N)$ in our
calculation.\footnote{The $\mu$-dependence in the above expressions is
absorbed into the five-point contact term~\cite{Filin:2013uma} whose
contributions to the pion production amplitude is not included in the
shown results.}
Furthermore,  the variables $x$, $y$ are defined as $x= k_1^2 / \mpi^2$, $y = \delta^2 / \mpi^2 $.
  The  permutations  $(1 \leftrightarrow 2)$  in the expressions above would result in   a symmetry factor of   four,  once the operators are  projected 
  onto the   partial wave  $^3P_1\to (^3S_1-^3D_1) s$ relevant for  the $pp\to d\pi^+$ reaction.
Further,  in order to obtain the  observables,  the operators above need  to  be convoluted with the  initial $NN$ and  final  deuteron wave functions.  The technical details of this procedure were discussed   in Ref.~\cite{Filin:2013uma} (see   appendix A).

 \section{Additional $N\Delta\to NN\pi$ operators in the coupled-channel  approach} 

The expressions for the  tree-level    diagrams involving  an initial or final state
$\Delta$ resonance, as  shown  in Fig.~\ref{fig:treeDelta}, read
\begin{eqnarray}
i M_{\rm dir\Delta a}  &=&
    \frac{\gpind \, (2m_N)}{\mN \fpi} \, T_1^a \,m_{\pi} \, (\vec S_1 \cdot \vec p)   (2\pi)^3   \delta(\vec p- {{\vec p}}^{\,\prime}),
    \nonumber
\\
i M_{\rm dir\Delta b}  &=&
    \frac{\gpind\, (2m_N)}{\mN \fpi} \, T_1^{\dagger a} \, m_{\pi} \,  (\vec S^{\,\dagger}_1 \cdot \vec p^{\,\prime})  (2\pi)^3   \delta(\vec p- {{\vec p}}^{\,\prime}),
    \nonumber
\\
i M_{\rm rescat\Delta a}&=&
    \frac{\gpind\, (2m_N)^2}{2 \fpi^3} \, i \epsilon^{bac} \tau_1^c T_2^b\, 
     \frac{ m_{\pi} }{2\omega(k)} \left(\frac1{P_1}+\frac1{P_{2\Delta}}\right)    (\vec S_2 \cdot \vec k),
    \nonumber
\\
i M_{\rm rescat\Delta b}&=&
    \frac{\gpind\, (2m_N)^2}{2 \fpi^3}\,   i \epsilon^{bac} \tau_1^c T_2^{\dagger b}\, 
     \frac{ m_{\pi} }{2\omega(k)} \left(\frac1{P_{1\Delta}}+\frac1{P_{2}}\right)
     (\vec S_2 \cdot \vec k),
    \label{Mdel}
\end{eqnarray}
where  
 $\vec S$ and ${\bf T}$
are the spin and isospin transition matrices,
normalized such that
\begin{eqnarray}
S_i S_j^\dagger = \frac{1}{3}\left( 2\delta_{ij}-i\epsilon_{ijk}\, \sigma_k\right),
\quad
T_i T_j^\dagger = \frac{1}{3}\left( 2\delta_{ij}-i\epsilon_{ijk}\, \tau_k\right)
\, ,  \quad i,j=1,2,3 .
%\nonumber
\end{eqnarray}
Furthermore, the TOPT propagators  read
\begin{eqnarray}
P_{1\Delta}&=&\sqrt{s}  - 2m_N -\delta - \frac{\vec p^{\,  2}}{2m_N} - \frac{\vec p^{\,\prime 2}}{2m_N}  - \omega(k) , \nonumber\\
 P_{2\Delta}&=&\sqrt{s}  - 2m_N-m_{\pi}-\delta - \frac{\vec p^{\,  2}}{2m_N} - \frac{\vec p^{\,\prime 2}}{2m_N}   - \omega(k).  
\end{eqnarray} 
Clearly, the operators \eqref{Mdel}   contribute
to the reaction amplitude of  $NN\to NN\pi$ only  when they are inserted as
a building block into
those of final-   and initial-state interaction diagrams which have an $N\Delta$  intermediate state like in
our coupled-channel
treatment  (CC$\chi$EFT$-\Delta$).
As already explained, to avoid double counting,  in this case the contributions of the diagrams in the second row  in Fig.~\ref{fig:allN2LO}  should not be included.


\begin{thebibliography}{99}
\bibitem{Hanhart:2003pg}
  C.~Hanhart,
  %``Meson production in nucleon-nucleon collisions close to the threshold,''
  Phys.\ Rept.\  {\bf 397} (2004) 155.
  % [hep-ph/0311341].
  %%CITATION = HEP-PH/0311341;%%

\bibitem{Baru:2013zpa}
  V.~Baru, C.~Hanhart and F.~Myhrer,
  %``Effective field theory calculations of $NN\to NN\pi$,''
  Int.\ J.\ Mod.\ Phys.\ E {\bf 23} (2014) 4,  1430004.
  % [arXiv:1310.3505 [nucl-th]].
  %%CITATION = ARXIV:1310.3505;%%

\bibitem{Hanhart:2000gp}
  C.~Hanhart, U.~van Kolck and G.~A.~Miller,
  %``Chiral three nucleon forces from p wave pion production,''
  Phys.\ Rev.\ Lett.\ \ {\bf 85},  (2000) 2905.
  %[nucl-th/0004033].
  %%CITATION = PRLTA,85,2905;%%


\bibitem{Baru:2009fm}
  V.~Baru, E.~Epelbaum, J.~Haidenbauer, C.~Hanhart, A.~E.~Kudryavtsev, V.~Lensky and U.-G.~Mei{\ss}ner,
  %``p-wave pion production from nucleon-nucleon collisions,''
  Phys.\ Rev.\ C\ {\bf 80},  (2009) 044003.
  %[arXiv:0907.3911 [nucl-th]].
  %%CITATION = PHRVA,C80,044003;%%

\bibitem{vanKolck:1994yi}
  U.~van Kolck,
  %``Few nucleon forces from chiral Lagrangians,''
  Phys.\ Rev.\ C {\bf 49},  (1994) 2932.
%  doi:10.1103/PhysRevC.49.2932
  %%CITATION = doi:10.1103/PhysRevC.49.2932;%%
  %356 citations counted in INSPIRE as of 27 Jan 2016


\bibitem{Epelbaum:2002vt}
  E.~Epelbaum, A.~Nogga, W.~Gloeckle, H.~Kamada, U.-G.~Mei{\ss}ner and H.~Witala,
  %``Three nucleon forces from chiral effective field theory,''
  Phys.\ Rev.\ C {\bf 66},  (2002) 064001.
  %doi:10.1103/PhysRevC.66.064001 [nucl-th/0208023].
  %%CITATION = doi:10.1103/PhysRevC.66.064001;%%
  %333 citations counted in INSPIRE as of 27 janv. 2016


\bibitem{Koltun:1965yk}
  D.~S.~Koltun and A.~Reitan,
  %``Production and absorption of S-wave pions at low energy by two nucleons,''
  Phys.\ Rev.\  {\bf 141} (1966) 1413.
  %%CITATION = PHRVA,141,1413;%%

\bibitem{meyer1992}
H.~O.~Meyer {\it et al.}, Nucl. Phys. A {\bf 539} (1992) 633.

\bibitem{Cohen:1995cc}
  T.~D.~Cohen, J.~L.~Friar, G.~A.~Miller and U.~van Kolck,
  %``The p p ---> p p pi0 reaction near threshold: A Chiral power counting approach,''
  Phys.\ Rev.\ C {\bf 53} (1996) 2661.
  % [nucl-th/9512036].
  %%CITATION = NUCL-TH/9512036;%%

\bibitem{Park:1995ku}
  B.~Y.~Park, F.~Myhrer, J.~R.~Morones, T.~Meissner and K.~Kubodera,
  %``Chiral perturbation approach to the p p ---> p p pi0 reaction near threshold,''
  Phys.\ Rev.\ C {\bf 53} (1996) 1519.
  % [nucl-th/9512023].
  %%CITATION = NUCL-TH/9512023;%%

\bibitem{Sato:1997ps}
  T.~Sato, T.~S.~H.~Lee, F.~Myhrer and K.~Kubodera,
  %``Chiral perturbation theory and the p p ---> p p pi0 reaction near threshold,''
  Phys.\ Rev.\ C {\bf 56} (1997) 1246.
  % [nucl-th/9704003].
  %%CITATION = NUCL-TH/9704003;%%

\bibitem{Hanhart:2002bu}
  C.~Hanhart and N.~Kaiser,
  %``Complete next-to-leading order calculation for pion production in nucleon-nucleon collisions at threshold,''
  Phys.\ Rev.\ C {\bf 66} (2002) 054005.
  % [nucl-th/0208050].
  %%CITATION = NUCL-TH/0208050;%%

\bibitem{Lensky:2005jc}
  V.~Lensky, V.~Baru, J.~Haidenbauer, C.~Hanhart, A.~E.~Kudryavtsev and U.-G.~Mei{\ss}ner,
  %``Towards a field theoretic understanding of NN ---> NN pi,''
  Eur.\ Phys.\ J.\ A {\bf 27} (2006) 37.
  % [nucl-th/0511054].
  %%CITATION = NUCL-TH/0511054;%%

\bibitem{Filin:2012za}
  A.~A.~Filin, V.~Baru, E.~Epelbaum, H.~Krebs, C.~Hanhart, A.~E.~Kudryavtsev and F.~Myhrer,
  %``Pion production in nucleon-nucleon collisions in chiral effective field theory: next-to-next-to-leading order contributions,''
  Phys.\ Rev.\ C {\bf 85} (2012) 054001.
  % [arXiv:1201.4331 [nucl-th]].
  %%CITATION = ARXIV:1201.4331;%%

\bibitem{Filin:2013uma}
  A.~A.~Filin, V.~Baru, E.~Epelbaum, H.~Krebs, C.~Hanhart and F.~Myhrer,
  %``Pion production in nucleon-nucleon collisions in chiral effective field theory with $\Delta(1232)$ degrees of freedom,''
  Phys.\ Rev.\ C {\bf 88} (2013)   064003.
  % [arXiv:1307.6187 [nucl-th]].
  %%CITATION = ARXIV:1307.6187;%%

\bibitem{Miller:2006tv}
  G.~A.~Miller, A.~K.~Opper and E.~J.~Stephenson,
  %``Charge symmetry breaking and QCD,''
  Ann.\ Rev.\ Nucl.\ Part.\ Sci.\  {\bf 56} (2006) 253.
  % [nucl-ex/0602021].
  %%CITATION = NUCL-EX/0602021;%%

\bibitem{Opper:2003sb}
  A.~K.~Opper {\it et al.},
  %``Charge symmetry breaking in n p ---> d pi0,''
  Phys.\ Rev.\ Lett.\  {\bf 91} (2003) 212302.
  % [nucl-ex/0306027].
  %%CITATION = NUCL-EX/0306027;%%

\bibitem{vanKolck:2000ip}
  U.~van Kolck, J.~A.~Niskanen and G.~A.~Miller,
  %``Charge symmetry violation in p n ---> d pi0 as a test of chiral effective field theory,''
  Phys.\ Lett.\ B {\bf 493} (2000) 65.
  % [nucl-th/0006042].
  %%CITATION = NUCL-TH/0006042;%%

\bibitem{Bolton:2009rq}
  D.~R.~Bolton and G.~A.~Miller,
  %``Charge Symmetry Breaking in the n p ---> d pi0 reaction,''
  Phys.\ Rev.\ C {\bf 81} (2010) 014001.
  % [arXiv:0907.0254 [nucl-th]].
  %%CITATION = ARXIV:0907.0254;%%

\bibitem{Filin:2009yh}
  A.~Filin, V.~Baru, E.~Epelbaum, J.~Haidenbauer, C.~Hanhart, A.~E.~Kudryavtsev and \\U.-G.~Mei{\ss}ner,
  %``Extraction of the strong neutron-proton mass difference from the charge symmetry breaking in pn ---> d pi0,''
  Phys.\ Lett.\ B {\bf 681} (2009) 423.
  % [arXiv:0907.4671 [nucl-th]].
  %%CITATION = ARXIV:0907.4671;%%

\bibitem{Strauch:2010rm}
  T.~Strauch {\it et al.},
  %``Precision determination of the dpi -> NN transition strength at threshold,''
  Phys.\ Rev.\ Lett.\  {\bf 104} (2010) 142503.
  % [arXiv:1003.4153 [nucl-ex]].
  %%CITATION = ARXIV:1003.4153;%%

\bibitem{Strauch:2010vu}
  T.~Strauch {\it et al.},
  %``Pionic deuterium,''
  Eur.\ Phys.\ J.\ A {\bf 47} (2011) 88.
  % [arXiv:1011.2415 [nucl-ex]].
  %%CITATION = ARXIV:1011.2415;%%

\bibitem{Weinberg:1992yk}
  S.~Weinberg,
  %``Three body interactions among nucleons and pions,''
  Phys.\ Lett.\ B {\bf 295} (1992) 114.
  % [hep-ph/9209257].
  %%CITATION = HEP-PH/9209257;%%

\bibitem{Hanhart:1997jd}
  C.~Hanhart, J.~Haidenbauer, M.~Hoffmann, U.-G.~Mei{\ss}ner and J.~Speth,
  %``The Reactions p p ---> p p pi0 and p p ---> d pi+ at threshold: The Role of the isoscalar pi N scattering amplitude,''
  Phys.\ Lett.\ B {\bf 424} (1998) 8.
  % [nucl-th/9707029].
  %%CITATION = NUCL-TH/9707029;%%

\bibitem{Dmitrasinovic:1999cu}
  V.~Dmitrasinovic, K.~Kubodera, F.~Myhrer and T.~Sato,
  %``A Next-to-next-to leading order p p ---> p p pi0 transition operator in chiral perturbation theory,''
  Phys.\ Lett.\ B {\bf 465} (1999) 43.
  % [nucl-th/9902048].
  %%CITATION = NUCL-TH/9902048;%%

\bibitem{Ando:2000ema}
  S.-i.~Ando, T.-S.~Park and D.~P.~Min,
  %``Threshold p p ---> p p pi0 up to one loop accuracy,''
  Phys.\ Lett.\ B {\bf 509} (2001) 253.
  % [nucl-th/0003004].
  %%CITATION = NUCL-TH/0003004;%%

\bibitem{Bernard:1998sz}
  V.~Bernard, N.~Kaiser and U.-G.~Mei{\ss}ner,
  %``Novel approach to pion and eta production in proton proton collisions,''
  Eur.\ Phys.\ J.\ A {\bf 4} (1999) 259.
  % [nucl-th/9806013].
  %%CITATION = NUCL-TH/9806013;%%

\bibitem{Kim:2007eh}
  Y.~Kim, T.~Sato, F.~Myhrer and K.~Kubodera,
  %``Two-pion-exchange contributions to the pp ---> pp pi0 reaction,''
  Phys.\ Lett.\ B {\bf 657},  (2007) 187.
%  doi:10.1016/j.physletb.2007.10.023
%  [arXiv:0704.1342 [nucl-th]].

\bibitem{Kim:2008qha}
  Y.~Kim, T.~Sato, F.~Myhrer and K.~Kubodera,
  %``Two-pion-exchange and other higher-order contributions to the pp ---> pp pi0 reaction,''
  Phys.\ Rev.\ C {\bf 80}, (2009) 015206.
 % doi:10.1103/PhysRevC.80.015206
 % [arXiv:0810.2774 [nucl-th]].


\bibitem{daRocha:1999dm}
  C.~da Rocha, G.~Miller and U.~van Kolck,
  %``The N N ---> N N pi+ reaction near threshold in a chiral power counting approach,''
  Phys.\ Rev.\ C {\bf 61} (2000) 034613.
  % [nucl-th/9904031].
  %%CITATION = NUCL-TH/9904031;%%

\bibitem{NNpiMenu}
  V.~Baru, J.~Haidenbauer, C.~Hanhart, A.~E.~Kudryavtsev, V.~Lensky and U.-G.~Mei{\ss}ner,
  %``Progress in NN ---> NN pi,''
 in {\it Proceedings of 11-th International
Conference on Meson-Nucleon Physics and the
Structure of the Nucleon (MENU 2007), J\" ulich, Germany},   eConfC\ {\bf 070910} 128 (2007);
  [arXiv:0711.2748 [nucl-th]].
  %%CITATION = ECONF,C070910,128;%%


  \bibitem{Niskanen:1978vm}
  J.~A.~Niskanen,
  %``The Differential Cross-Section and Polarization in p p --> d pi+,''
  Nucl.\ Phys.\ A {\bf 298}, (1978) 417.

\bibitem{Hanhart:1998za}
  C.~Hanhart, J.~Haidenbauer, O.~Krehl and J.~Speth,
  %``Role of the Delta isobar in the reaction N N ---> N N pi near threshold,''
  Phys.\ Lett.\ B {\bf 444} (1998) 25.
%  doi:10.1016/S0370-2693(98)01385-9 [nucl-th/9808020].

\bibitem{Hemmert:1997ye}
  T.~R.~Hemmert, B.~R.~Holstein and J.~Kambor,
  %``Chiral Lagrangians and delta(1232) interactions: Formalism,''
  J.\ Phys.\ G {\bf 24},  (1998) 1831.
%  doi:10.1088/0954-3899/24/10/003 [hep-ph/9712496].
  %%CITATION = doi:10.1088/0954-3899/24/10/003;%%
  %238 citations counted in INSPIRE as of 27 janv. 2016

\bibitem{Pascalutsa:2002pi}
  V.~Pascalutsa and D.~R.~Phillips,
  %``Effective theory of the delta(1232) in Compton scattering off the nucleon,''
  Phys.\ Rev.\ C {\bf 67},  (2003) 055202.
%  doi:10.1103/PhysRevC.67.055202 [nucl-th/0212024].
  %%CITATION = doi:10.1103/PhysRevC.67.055202;%%
  %139 citations counted in INSPIRE as of 27 janv. 2016

\bibitem{Deltuva:2003wm}
  A.~Deltuva, R.~Machleidt and P.~U.~Sauer,
  %``Realistic two baryon potential coupling two nucleon and nucleon delta isobar states: Fit and applications to three nucleon system,''
  Phys.\ Rev.\ C {\bf 68},  (2003) 024005.
%  doi:10.1103/PhysRevC.68.024005
  %%CITATION = doi:10.1103/PhysRevC.68.024005;%%



\bibitem{CCF}
  J.~Haidenbauer, K.~Holinde and M.~B.~Johnson,
  %``A Coupled channel potential for nucleons and deltas,''
  Phys.\ Rev.\ C {\bf 48},    (1993) 2190.
  %%CITATION = PHRVA,C48,2190;%%

  \bibitem{Baru:2007wf}
  V.~Baru, J.~Haidenbauer, C.~Hanhart, A.~E.~Kudryavtsev, V.~Lensky and U.-G.~Mei{\ss}ner,
  %``Role of the Delta(1232) in pion-deuteron scattering at threshold within chiral effective field theory,''
  Phys.\ Lett.\ B {\bf 659},  (2008) 184.
%  doi:10.1016/j.physletb.2007.10.063
 % [arXiv:0706.4023 [nucl-th]].
  %%CITATION = doi:10.1016/j.physletb.2007.10.063;%%

  \bibitem{Baru:2011bw}
  V.~Baru, C.~Hanhart, M.~Hoferichter, B.~Kubis, A.~Nogga and D.~R.~Phillips,
  %``Precision calculation of threshold pi^- d scattering, pi N scattering lengths, and the GMO sum rule,''
  Nucl.\ Phys.\ A {\bf 872},  (2011) 69.
%  doi:10.1016/j.nuclphysa.2011.09.015
  %[arXiv:1107.5509 [nucl-th]].
  %%CITATION = doi:10.1016/j.nuclphysa.2011.09.015;%%

  \bibitem{Baru:2010xn}
  V.~Baru, C.~Hanhart, M.~Hoferichter, B.~Kubis, A.~Nogga and D.~R.~Phillips,
  %``Precision calculation of the $\pi^{-}$ deuteron scattering length and its impact on threshold $\pi$ N scattering,''
  Phys.\ Lett.\ B {\bf 694},  (2011) 473.
  %doi:10.1016/j.physletb.2010.10.028
  %[arXiv:1003.4444 [nucl-th]].

\bibitem{Machleidt:2000ge} % cdbonn 2000 potential
  R.~Machleidt,
  %``The High precision, charge dependent Bonn nucleon-nucleon potential (CD-Bonn),''
  Phys.\ Rev.\ C {\bf 63} (2001) 024001.
  % [nucl-th/0006014].
  %%CITATION = NUCL-TH/0006014;%%

\bibitem{Stoks:1994wp} % of potential
  V.~G.~J.~Stoks, R.~A.~M.~Klomp, C.~P.~F.~Terheggen and J.~J.~de Swart,
  %``Construction of high quality N N potential models,''
  Phys.\ Rev.\ C {\bf 49} (1994) 2950.
  % [nucl-th/9406039].
  %%CITATION = NUCL-TH/9406039;%%

\bibitem{Wiringa:1994wb} %AV18 potential
  R.~B.~Wiringa, V.~G.~J.~Stoks and R.~Schiavilla,
  %``An Accurate nucleon-nucleon potential with charge independence breaking,''
  Phys.\ Rev.\ C {\bf 51} (1995) 38.
  % [nucl-th/9408016].
  %%CITATION = NUCL-TH/9408016;%%

\bibitem{Krebs:2007rh}
  H.~Krebs, E.~Epelbaum and U.-G.~Mei{\ss}ner,
  %``Nuclear forces with Delta-excitations up to next-to-next-to-leading order. I. Peripheral nucleon-nucleon waves,''
  Eur.\ Phys.\ J.\ A {\bf 32}, (2007) 127.
%  doi:10.1140/epja/i2007-10372-y
 % [nucl-th/0703087].



\bibitem{Fettes:1998ud}
  N.~Fettes, U.-G.~Mei{\ss}ner and S.~Steininger,
  %``Pion - nucleon scattering in chiral perturbation theory. 1. Isospin symmetric case,''
  Nucl.\ Phys.\ A {\bf 640},  (1998) 199.
%  doi:10.1016/S0375-9474(98)00452-7 [hep-ph/9803266].



\bibitem{Krebs:2012yv}
  H.~Krebs, A.~Gasparyan and E.~Epelbaum,
  %``Chiral three-nucleon force at N^4LO I: Longest-range contributions,''
  Phys.\ Rev.\ C {\bf 85}, (2012) 054006.
 % [arXiv:1203.0067 [nucl-th]].
  %%CITATION = ARXIV:1203.0067;%%
  %31 citations counted in INSPIRE as of 11 Oct 2014


%\cite{Fettes:2000xg}
\bibitem{Fettes:2000xg}
  N.~Fettes and U.-G.~Mei{\ss}ner,
  %``Pion nucleon scattering in chiral perturbation theory. 2.: Fourth order calculation,''
  Nucl.\ Phys.\ A {\bf 676},  (2000) 311.
%  [hep-ph/0002162].
  %%CITATION = HEP-PH/0002162;%%
  %126 citations counted in INSPIRE as of 12 Oct 2014

%\cite{Alarcon:2012kn}
\bibitem{Alarcon:2012kn}
  J.~M.~Alarcon, J.~Martin Camalich and J.~A.~Oller,
  %``Improved description of the $\pi N$-scattering phenomenology in covariant baryon chiral perturbation theory,''
  Annals Phys.\  {\bf 336}, (2013) 413.
%  [arXiv:1210.4450 [hep-ph]].
  %%CITATION = ARXIV:1210.4450;%%
  %27 citations counted in INSPIRE as of 12 Oct 2014

%\cite{Chen:2012nx}
\bibitem{Chen:2012nx}
  Y.~H.~Chen, D.~L.~Yao and H.~Q.~Zheng,
  %``Analyses of pion-nucleon elastic scattering amplitudes up to $O(p^4)$ in extended-on-mass-shell subtraction scheme,''
  Phys.\ Rev.\ D {\bf 87},  (2013) 054019.
%  [arXiv:1212.1893 [hep-ph]].
  %%CITATION = ARXIV:1212.1893;%%
  %10 citations counted in INSPIRE as of 12 Oct 2014

%\cite{Wendt:2014lja}
\bibitem{Wendt:2014lja}
  K.~A.~Wendt, B.~D.~Carlsson and A.~Ekstr\"om,
  %``Uncertainty Quantification of the Pion-Nucleon Low-Energy Coupling Constants up to Fourth Order in Chiral Perturbation Theory,''
  arXiv:1410.0646 [nucl-th].
  %%CITATION = ARXIV:1410.0646;%%

\bibitem{Siemens:2016hdi} 
  D.~Siemens, V.~Bernard, E.~Epelbaum, A.~Gasparyan, H.~Krebs and U.-G.~Mei\ss ner,
  %``Elastic pion-nucleon scattering in chiral perturbation theory: A fresh look,''
  arXiv:1602.02640 [nucl-th].

\bibitem{Epelbaum:2014efa}
  E.~Epelbaum, H.~Krebs and U.-G.~Mei{\ss}ner,
  %``Improved chiral nucleon-nucleon potential up to next-to-next-to-next-to-leading order,''
  Eur.\ Phys.\ J.\ A {\bf 51} (2015) 5,  53.
  % [arXiv:1412.0142 [nucl-th]].
  %%CITATION = ARXIV:1412.0142;%%

\bibitem{Hoferichter:2015dsa}
  M.~Hoferichter, J.~Ruiz de Elvira, B.~Kubis and U.-G.~Mei{\ss}ner,
  %``High-Precision Determination of the Pion-Nucleon Term from Roy-Steiner Equations,''
  Phys.\ Rev.\ Lett.\  {\bf 115}, (2015) 092301.
%  doi:10.1103/PhysRevLett.115.092301  [arXiv:1506.04142 [hep-ph]].
  %%CITATION = doi:10.1103/PhysRevLett.115.092301;%%
  %19 citations counted in INSPIRE as of 27 janv. 2016

\bibitem{Hoferichter:2015tha}
  M.~Hoferichter, J.~Ruiz de Elvira, B.~Kubis and U.-G.~Mei{\ss}ner,
  %``Matching pion-nucleon Roy-Steiner equations to chiral perturbation theory,''
  Phys.\ Rev.\ Lett.\  {\bf 115},   (2015) 192301.
%  doi:10.1103/PhysRevLett.115.192301  [arXiv:1507.07552 [nucl-th]].
  %%CITATION = doi:10.1103/PhysRevLett.115.192301;%%
  %5 citations counted in INSPIRE as of 27 janv. 2016

\bibitem{Buettiker:1999ap}
  P.~B\"uttiker and U.-G.~Mei{\ss}ner,
  %``Pion nucleon scattering inside the Mandelstam triangle,''
  Nucl.\ Phys.\ A {\bf 668}, (2000) 97.
%  doi:10.1016/S0375-9474(99)00813-1
%  [hep-ph/9908247].

\bibitem{Bernard:1996gq}
  V.~Bernard, N.~Kaiser and U.-G.~Mei{\ss}ner,
  %``Aspects of chiral pion - nucleon physics,''
  Nucl.\ Phys.\ A {\bf 615}, (1997) 483.
%  doi:10.1016/S0375-9474(97)00021-3  [hep-ph/9611253].
  %%CITATION = doi:10.1016/S0375-9474(97)00021-3;%%
  %219 citations counted in INSPIRE as of 27 janv. 2016


  \bibitem{Epelbaum:2007sq}
  E.~Epelbaum, H.~Krebs and U.-G.~Mei{\ss}ner,
  %``Delta-excitations and the three-nucleon force,''
  Nucl.\ Phys.\ A {\bf 806}, (2008) 65.
%  doi:10.1016/j.nuclphysa.2008.02.305  [arXiv:0712.1969 [nucl-th]].




\bibitem{Epelbaum:2014sza}
  E.~Epelbaum, H.~Krebs and U.-G.~Mei{\ss}ner,
  %``Precision nucleon-nucleon potential at fifth order in the chiral expansion,''
  Phys.\ Rev.\ Lett.\  {\bf 115} (2015) 12,  122301.
  % [arXiv:1412.4623 [nucl-th]].
  %%CITATION = ARXIV:1412.4623;%%

\bibitem{Entem:2003ft}
  D.~R.~Entem and R.~Machleidt,
  %``Accurate charge dependent nucleon nucleon potential at fourth order of chiral perturbation theory,''
  Phys.\ Rev.\ C {\bf 68}, (2003) 041001.
%  doi:10.1103/PhysRevC.68.041001  [nucl-th/0304018].
  %%CITATION = doi:10.1103/PhysRevC.68.041001;%%
  %777 citations counted in INSPIRE as of 27 Jan 2016

\bibitem{Epelbaum:2004fk}
  E.~Epelbaum, W.~Gl\"ockle and U.-G.~Mei{\ss}ner,
  %``The Two-nucleon system at next-to-next-to-next-to-leading order,''
  Nucl.\ Phys.\ A {\bf 747}, (2005) 362.
%  doi:10.1016/j.nuclphysa.2004.09.107  [nucl-th/0405048].
  %%CITATION = doi:10.1016/j.nuclphysa.2004.09.107;%%
  %419 citations counted in INSPIRE as of 27 janv. 2016

\bibitem{Filin_CDtalk}
  A.~A.~Filin, V.~Baru, E.~Epelbaum, H.~Krebs, C.~Hanhart, and F.~Myhrer,
  %``Pion production in nucleon-nucleon collisions in chiral effective field theory: next-to-next-to-leading order contributions,''
   in {\it Proceedings of 8-th International Workshop  on Chiral Dynamics   (CD 2015), Pisa, Italy},  to be published.
 %   eConfC\ {\bf 070910} 128 (2007);[arXiv:0711.2748 [nucl-th]].
  % [arXiv:1201.4331 [nucl-th]].
  %%CITATION = ARXIV:1201.4331;%%

\end{thebibliography}
\end{document}